\documentclass[aps,rmp,twocolumn,groupedaddress,floatfix,showpacs]{revtex4}

\usepackage[]{graphicx}
\usepackage{fancyhdr}
\usepackage{bm}
\newcommand{\mrm}[1]{\mathrm{#1}}
\newcommand{\mbf}[1]{\mathbf{#1}}
\newcommand{\gprime}{\mbf{g}^\prime}

\begin{document}
\urldef{\ashcopy}\url{http://link.aps.org/abstract/PRL/v30/p139}

\title{Momentum of an electromagnetic wave in dielectric media}


\author{Robert N. C. Pfeifer}
\email{pfeifer@physics.uq.edu.au}
\homepage{http://www.physics.uq.edu.au/people/pfeifer/}
\author{Timo A. Nieminen}
\email{timo@physics.uq.edu.au}
\author{Norman R. Heckenberg}
\email{heckenberg@physics.uq.edu.au}
\author{Halina Rubinsztein-Dunlop}
\email{halina@physics.uq.edu.au}
\affiliation{Centre for Biophotonics and Laser Science, School of Physical Sciences, The University of Queensland, Brisbane, QLD 4072, Australia}


\date{May 8, 2007} 
\begin{abstract}
Almost a hundred years ago, two different expressions were proposed for the energy--momentum tensor of an electromagnetic wave in a dielectric. Minkowski's tensor predicted an increase in the linear momentum of the wave on entering a dielectric medium, whereas Abraham's tensor predicted its decrease. Theoretical arguments were advanced in favour of both sides, and experiments proved incapable of distinguishing between the two. Yet more forms were proposed, each with their advocates who considered the form that they were proposing to be the one true tensor. This paper reviews the debate and its eventual conclusion: that no electromagnetic wave energy--momentum tensor is complete on its own. When the appropriate accompanying energy--momentum tensor for the material medium is also considered, experimental predictions of all the various proposed tensors will always be the same, and the preferred form is therefore effectively a matter of personal choice.
\end{abstract}

\pacs{03.50.De, 41.20.Jb\\\ \\Published as: Reviews of Modern Physics 79(4), 1197-1216 (2007)}


\maketitle

\tableofcontents

\section{Introduction}

The correct form of the energy--momentum tensor of an electromagnetic wave, and hence the momentum of an electromagnetic wave in a dielectric medium, has been debated for almost a century. Two different forms of the energy--momentum tensor were originally proposed by \textcite{minkowski,minkowski2}
and
\textcite{abraham,abraham2}, though more have been added in later years \cite{penfield,grot,degroot2a,degroot4,degroot,marx,peierls,livens}.
The question regularly attracts experimental interest, as on initial inspection it may appear that some of the different tensors give rise to distinct and potentially useful physical consequences.

We discuss why this is not the case, reviewing the key experiments of
\textcite{jones}, \textcite{ashkin}, and \textcite{walker},
and theoretical work including that of 
\textcite{penfield}, \textcite{degroot}, \textcite{gordon}, \textcite{mikura}, \textcite{kranys,kranys3,kranys2}, and \textcite{maugin},
the latter four authors demonstrating explicitly and directly the equivalence of a number of different energy--momentum tensors.

More recent work on the subject may largely be divided into three camps. First, there are those concerned with practical applications, focussing primarily on which expression is most useful in a given circumstance. Second, there are those extending the theoretical work into new domains. Frequently such papers are hampered by the fragmentary nature of existing literature on the topic and their conclusions may be weakened as a result. Third, there are those who are unconvinced that the controversy has been resolved, perhaps unsurprising as certain key publications remain relatively obscure.
It is clear that broad interest still exists in the subject, and from the widespread prevalence of work in the second and third categories, that the field stands to benefit greatly from a coherent and comprehensive review.

We therefore hope this paper will increase awareness that the controversy has been resolved, and that predictions regarding measurable behaviours will always be independent of the electromagnetic energy--momentum tensor chosen, provided the accompanying material tensor is also taken into account.

In particular, renewed interest in the controversy has been stimulated by the advent of optical trapping and manipulation of microparticles in fluid media \cite{ashkin1986,lang2003}.
We wish to lay to rest any concerns that predictions of observable effects in optical tweezers might depend on the division of momentum between field and medium,
and to provide an introduction to the subject of the energy--momentum tensor in a dielectric medium. We hope that by reading this paper and some of its references, those new to the field will gain a comprehensive understanding of this interesting subject.

\section{A Sound Foundation}

The Abraham--Minkowski controversy debates the correct expression for electromagnetic momentum within a material medium, and by implication the form of the four-dimensional energy--momentum tensor. Before we debate the relative merits of the different expressions, we must first satisfy ourselves that it is possible to construct a well-defined energy--momentum tensor. After all, on a microscopic level the transfer of energy and momentum consists of the absorption and emission of photons as they interact with the atoms of the medium, a process which is both stochastic and quantum-mechanical in nature.

Fortunately, the Abraham--Minkowski controversy is conventionally formulated wholly within the realms of classical continuum electrodynamics \cite{nelson}, in which the fluctuations due to the inherent unpredictability of the microscopic realm become negligible. Specifically, the properties of the medium must change negligibly over unit cells which are nevertheless large in comparison with the mean volume per atom. Discontinuous inter-medium boundaries are permitted provided they are sharp on the scale of these unit cells. The transition from stochastic behaviour on the microscopic scale to the Maxwell equations of the continuum, and hence by implication a domain in which both the refractive index and the energy--momentum tensor are well-defined and piecewise smooth is comprehensively detailed by \textcite{degroot}.

A number of authors do extend the problem into the microscopic domain \cite{garrison,hensley,haugan,loudon3}. Such work must of course be evaluated on its own merits, and on its consistency with the results of experiment \cite{jones2,campbell,gibson}.

\section{The Early Years\label{earlyyears}}

The first author to propose an expression for the energy--momentum tensor of an electromagnetic wave in a dielectric medium was
\textcite{minkowski}.
His expression corresponds to an electromagnetic wave momentum density of $\mathbf{D}\times\mathbf{B}$, where $\mathbf{D}$ is the electric displacement field and $\mathbf{B}$ is the magnetic flux density. The total momentum 
of a propagating electromagnetic wave therefore increases upon entering a dielectric medium, from $p$ in free space, to $np$, where $n$ is the refractive index of the medium, and we are neglecting dispersion. An accessible derivation of the Minkowski energy--momentum tensor may be found in
\textcite{moller2}.

The electromagnetic energy--momentum tensor of Minkowski was not diagonally symmetric, and this drew considerable criticism as it was held to be incompatible with the conservation of angular momentum. 
\textcite{abraham,abraham2} therefore developed an alternative, symmetric, tensor, for which the electromagnetic wave momentum density was instead $(1/c^2)(\mathbf{E}\times\mathbf{H})$, with $\mathbf{E}$ representing the electric field and $\mathbf{H}$ the magnetic field. The momentum
of an electromagnetic wave entering a dielectric medium then falls to $p/n$. Under the Abraham tensor, a photon therefore carries less momentum within a medium than in the Minkowski case.

However, both of these tensors are incomplete in isolation \cite{penfield,degroot}. 
Momentum may also be carried by interactions occurring within the material medium, and between the medium and the electromagnetic field. 
This is most obvious for the Abraham tensor, as the material tensor must be considered in order to obtain the correct rate of transfer of momentum to a reflective object suspended within a dielectric medium, as in the experiments of \textcite{jones} and \textcite{jones2}.
However, increasing the photon momentum from $p$ to $np$ in accordance with the Minkowski tensor yields the correct result directly,
and hence many authors neglect to observe that it, too, is associated with a corresponding material tensor.
Any complete, thermodynamically closed system must conserve energy and momentum, and it is only by taking into account the corresponding material tensor that the Minkowski formulation can be reconciled with conservation of total angular momentum \cite{brevik4}.
The behaviour of a transparent object is less straightforward, with the object acquiring a fixed velocity away from the source while exposed to the electromagnetic field. The existing literature lacks a single consistent mathematical approach capable of dealing with wholly or partially reflective and transparent objects immersed within a dielectric medium. We address this lack in Sec.~\ref{totememt}.

When the force exerted by an electromagnetic wave in a dielectric medium is evaluated without recourse to the material counterpart of the Minkowski tensor, then the two tensors will predict different results. The force density predicted by the Abraham tensor pair is smaller by
\begin{equation}
\mathbf{f_{Abr}}=\frac{\varepsilon_\mrm{r}\mu_\mrm{r} -1}{c^2}\frac{\partial \mathbf{S}}{\partial t}\label{eq:abrterm},
\end{equation}
which is known historically as the Abraham force. Here, $\mathbf{S}$ is the real instantaneous Poynting vector $\mathbf{E}\times\mathbf{H}$, and $\varepsilon_\mrm{r}$ and $\mu_\mrm{r}$ are the relative permittivity 
and permeability 
of the medium respectively. When the material counterpart to the Minkowski electromagnetic energy--momentum tensor is also employed, as described in Sec.~\ref{sec:atotal}, then the Minkowski tensor pair also gives rise to the smaller value.

In the early part of the 20th Century the importance of the material energy--momentum tensor was not fully appreciated, and vigorous debate ensued regarding the relative merits of the Abraham and Minkowski electromagnetic energy--momentum tensors. We shall discuss more fully how the requirement for the material counterpart to the Minkowski tensor arises from conservation of total linear and angular momentum in Sec.~\ref{proofsequiv}.

Meanwhile, it is worth noting
that in some sense ``the momentum of a photon in a dielectric medium'' is an abstract concept, as it is impossible to construct an experiment to directly measure this momentum. Instead, one can only measure the total energy or momentum transferred to a detecting apparatus in a given set of
circumstances. This may include contributions both from the electromagnetic momentum and from the associated material energy--momentum tensor. One may therefore ask whether 
it is possible or not to devise an experiment
which distinguishes between the Abraham and the Minkowski tensors.

\section{Experiments\label{expts}}

\subsection{Jones and Richards\label{expts:jones}}

The first experiment to
measure the radiation pressure due to light in a refracting medium was conducted by
\textcite{jones3} at the University of Aberdeen.
Jones sought to verify the prediction that when a mirror was immersed in a dielectric medium, the radiation pressure exerted on the mirror would be proportional to the refractive index of the medium. Later refinement of this experiment led to a paper by 
\textcite{jones}
verifying this effect to an accuracy of $\sigma=\pm 1.2\%$. 

\begin{figure}
\includegraphics[width=8.5cm]{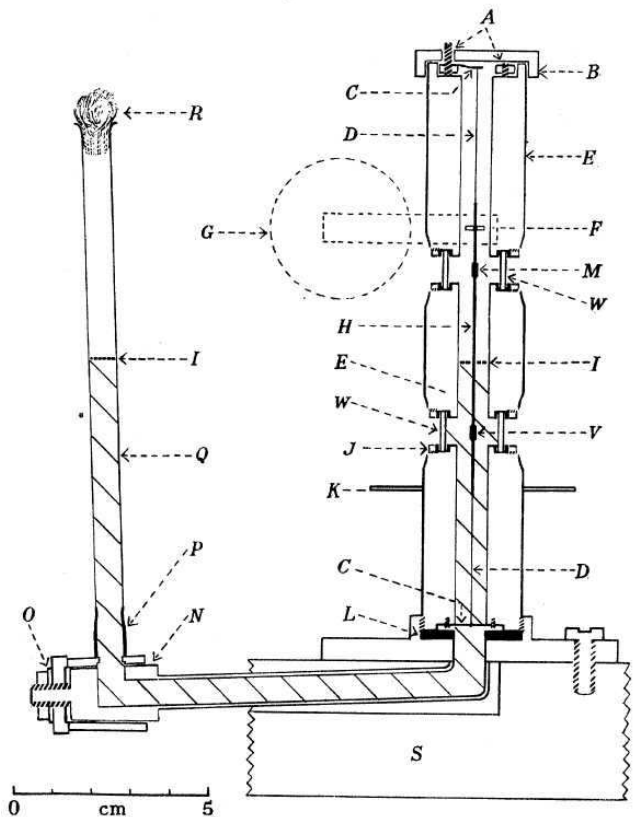}
\caption{The suspension and its container. \textit{A}, screws to tauten suspension; \textit{B}, brass draft cover; \textit{C}, phosphor bronze strip 0.005 cm thick; \textit{D}, Johnson Matthey gold alloy suspension 0.005 $\times$ 0.0005 cm; \textit{E}, brass tube 0.8 cm internal diam.; \textit{F}, magnet 0.2 cm long; \textit{G}, electromagnet; \textit{H}, 48 s.w.g. copper wire; \textit{I}, liquid levels; \textit{J}, screwed brass ring and lead washers; \textit{K}, support for mirrors; \textit{L}, lead washer; \textit{M}, silvered glass mirror 0.2 $\times$ 0.5 $\times$ 0.02 cm; \textit{N}, brass swivel joint; \textit{O}, brass nut and washers; \textit{P}, copper/glass seal; \textit{G}[\emph{sic}], glass tube; \textit{R}, cotton-wool plug; \textit{S}, triangular cast iron base; \textit{V}, rhodium-plated silver vane 0.2 $\times$ 0.5 $\times$ 0.01 cm; \textit{W}, circular glass windows 0.1 cm thick.
Reprinted figure with permission, from \textcite{jones}, \emph{The Pressure of Radiation in a Refracting Medium}, Fig.~1, published by and \copyright (1954) The Royal Society.\label{fig:jonesexpt}}
\end{figure}

The experimental set-up (shown in Fig.~\ref{fig:jonesexpt}) consisted of a pair of mirrors strung on a vertical wire and tethered at each end by a gold alloy torsion fibre.
The lower mirror (vane \textit{V}) was subjected to an asymmetric illumination, giving rise to a torque about the axis of suspension and causing both lower (\textit{V}) and upper (\textit{M}) mirrors to rotate until this torque was neutralised by the torsion fibre. By measuring the angle of the upper mirror, the authors determined the magnitude of the torque, and hence the rate of momentum transfer to the lower mirror. By immersing the lower mirror in a number of liquids of varying refractive index and observing the resultant deflections, the momentum transferred was found to scale with the refractive index, and the accuracy obtained was sufficient to show that this dependency was upon the phase refractive index ($n_\phi$), and not the group refractive index ($n_\mathrm{g}$) of the medium. This finding
agreed with theoretical expectations, as the behaviour examined was concerned with variations in the electric field at a point, rather than the rate of propagation through the medium of a modulation in the field.

In the theoretical introduction to the 1954 paper, Jones discussed whether the choice of expression for the momentum density of the electromagnetic wave was dependent either upon $\mathbf{D}\times\mathbf{B}$ or $\mathbf{E}\times\mathbf{H}$,
though without direct reference to the work of Minkowski or Abraham. He observed that the former followed naturally from the expression for the Maxwellian stresses on a body within a radiation field, whereas the latter could be suggested on relativistic grounds
\cite[see][]{pauli,pauli2} but was also accompanied by a corresponding stress in the surrounding medium of $(1/c)(k\mu-1)(\mathbf{E}\times\mathbf{H})$ (in Gaussian units),
where $k$ and $\mu$ stand for the dielectric constant and magnetic permeability of the medium respectively, and the expression is equivalent to the Abraham term (\ref{eq:abrterm}). He commented that ``it does not seem possible to devise a practicable experiment to distinguish between the two cases''.
\textcite{brevik3}
subsequently argued that when the linear momentum flux densities due to the Abraham electromagnetic tensor and the corresponding material stress tensor are combined, the result is the same as that given by the Minkowski tensor.
He thus interpreted the experiment of Jones and Richards
as providing more direct support for the Minkowski formulation.

Although in principle Jones and Richards are correct, and the Abraham electromagnetic momentum is accompanied by an induced material momentum, they are wrong to identify this momentum flux with the Abraham term. We shall examine the Abraham material momentum in greater detail in Sec.~\ref{sec:atotal} and return to this experiment in Sec.~\ref{ex:mirror}.

\subsection{Ashkin and Dziedzic\label{expts:ashkin}}

The next significant experiment conducted was that of
\textcite{ashkin},
in response to a theoretical paper by
\textcite{burt}.
In this paper, Burt and Peierls argued that when a laser beam passed from air into a dielectric fluid of greater refractive index, transfer of momentum to the fluid would cause the surface to either bulge outwards, if the Minkowski tensor was correct, or be depressed, if the Abraham tensor was correct. However, this conclusion was based on the mistaken assumption that the momentum of the electromagnetic wave would rapidly move ahead of any momentum associated with the material medium, allowing them to neglect the effects of the material tensor in the Abraham case.
In fact, the field-related material component of the momentum constitutes an excitation induced in the medium by the leading edge of the electromagnetic wave and eliminated by the trailing edge. The excitation of the material medium therefore propagates at the same velocity as the electromagnetic wave.

\begin{figure}
\includegraphics[width=\columnwidth]{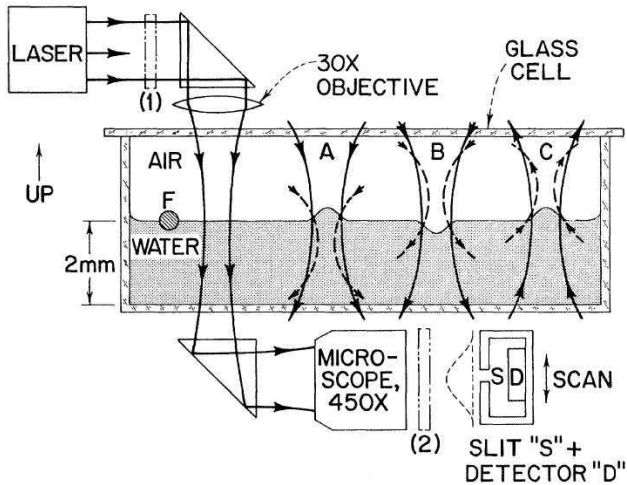}
\caption{Basic apparatus: \textit{A}, Beam shapes for low power (solid curve) and high power (dashed curve) for positive surface lens; \textit{B}, shapes (for low and high power) for a negative surface lens. For \textit{A} and \textit{B} the beam is incident from above. \textit{C}, beam shapes for low and high power for a positive surface lens with the beam incident from below. \textit{F} is a 2-4 $\mu\textrm{m}$ glass fibre used to identify the surface of the water under the microscope. A beam attenuator was placed either in position (1) or (2) depending on the power level desired within the glass cell.
Reprinted figure with permission, from
\textcite{ashkin}.
\ashcopy
\label{fig:ashkinexpt}}
\end{figure}

Ashkin and Dziedzic performed this experiment, passing a laser beam vertically through a glass cell containing air and water (Fig.~\ref{fig:ashkinexpt}).
Bulging or depression of the liquid surface would cause a lensing effect, altering the radial profile of the laser beam. By studying the beam profile as it emerged from the cell, they were able to determine that the surface of the liquid was caused to bulge outwards. In their paper they acknowledged work by \textcite{gordon}, at that time still unpublished,
which showed that this result does not invalidate the Abraham tensor, as the predictions of the two tensors are reconciled when the material counterpart to the Abraham tensor is properly taken into account.

\subsection{James; Walker, Lahoz and Walker\label{expts:walker}}

In 1968 James performed an experiment
to directly measure the Abraham force \cite{james,james2}. 
Two ferrite toroids, axially aligned with one another, were connected to a piezo-electric transducer. The toroids were subjected to time-varying electric and magnetic fields. Both Minkowski's and Abraham's energy--momentum tensors predicted a net torque on the toroids, but with differing magnitudes due to the existence of the Abraham term, Eq.~(\ref{eq:abrterm}). With the varying electric fields arranged in antiphase, the two toroids exerted torques in opposite directions upon the piezo-electric transducer, and by measuring the magnitude of this torque, James was able to experimentally demonstrate the existence of the Abraham force.

\begin{figure}
\includegraphics[width=\columnwidth]{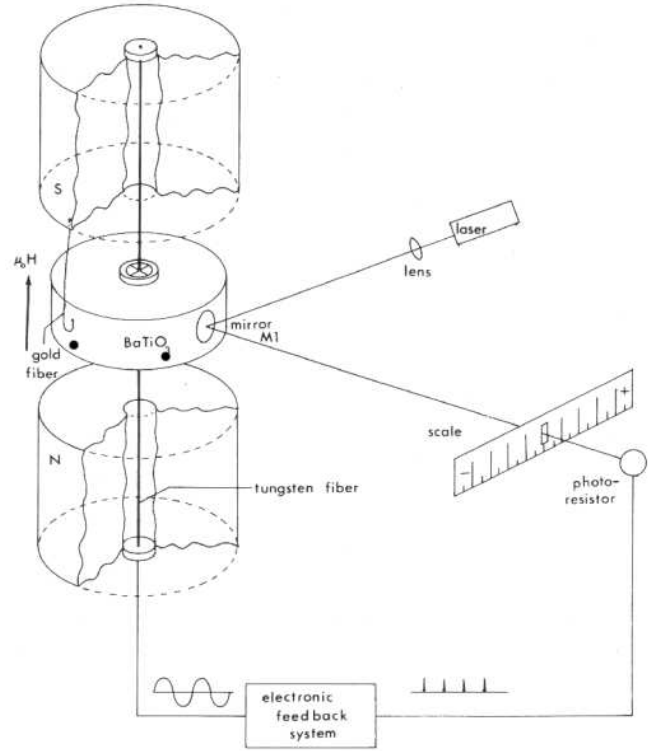}
\caption{Diagram of the experimental apparatus.
Reprinted figure with permission, from 
\textcite{walker}.
\copyright (1975) National Research Council, Canada.\label{fig:walkerexpt}}
\end{figure}

This result went largely unnoticed until the experiment of
\textcite{walker}.
This team measured the torque exerted on a disc of barium titanate which was suspended on a torsion fibre in a constant axial magnetic field, and subjected to a time-varying radial electric field (Fig.~\ref{fig:walkerexpt}).
The disc behaved as a torsion pendulum and its period of oscillation was measured by reflecting a laser beam off a mirror attached to the disc. To maximise the response of the system, the time period of the electric field was synchronised with the oscillations of the disc. Following an initial displacement, the oscillations of the disc settled exponentially towards a constant period, supporting the existence of the Abraham force.
The experiment was also later
modified to confirm the effects of a time-varying magnetic field \cite{walker2}.

As the conservation arguments leading to demonstration of the material counterpart of the Minkowski tensor were not widely recognised at this time, this experiment was taken as providing support for the Abraham tensor over that of Minkowski \cite{brevik3}.
However, 
\textcite{israel}
and later, tangentially,
\textcite{obukhov}
have demonstrated that for a suitable combination of electromagnetic and material energy--momentum tensors the observed results are also compatible with the Minkowski case, and hence these experiments cannot discriminate between the two.


\subsection{Jones and Leslie}

In 1977, Jones and Leslie repeated the experiment of 1954, taking advantage of recent advances in technology. A laser was employed as a light source instead of a tungsten lamp, allowing the refractive index of the liquid to be stated with greater precision as only one wavelength of light was involved; a superior mirror was employed, reducing heating of the dielectric liquids and hence convection forces in the fluid; and the torsion apparatus was restored to the neutral position by means of an external electromagnet. The current in this electromagnet related to the radiation pressure in a highly linear manner, whereas in the previous apparatus, the authors estimate that nonlinearity in the measured deflection of the secondary mirror may have been as large as $1\%$. 

Through these modifications to the experiment, Jones and Leslie obtained final results with a standard deviation of only $0.05\%$ \cite{jones2}. In the accompanying theoretical paper \cite{jones4}, Jones discussed whether this corresponded to a photon momentum of $n_\phi p$ in accordance with the Minkowski tensor, or $p/n_\mathrm{g}$ in accordance with the Abraham tensor (where $p$ is the momentum of the photon in free space, and $n_\phi$ and $n_g$ are respectively the phase and group refractive indices). Once again, he concluded that both were acceptable provided that in the Abraham case the total momentum $n_\phi p$ was divided into an electromagnetic component $p/n_\phi$, and a mechanical component, which was expressed this time as $n_\phi p[1-1/(n_\phi n_\mathrm{g})]$.

\section{Hypothetical Constraints and Alternate Theories\label{arguments}}

Numerous attempts have been made to find additional constraints by which one or another of the Abraham and Minkowski energy momentum tensors might be proven invalid, the earliest being the suggestion that the Minkowski tensor failed to conserve angular momentum. This led first
\textcite{einstein,einstein2}
and then
\textcite{abraham,abraham2}
to attempt to develop symmetric forms of the electromagnetic energy--momentum tensor. The Einstein--Laub tensor enjoyed only limited attention as it does not purport to be valid outside the rest frame of the dielectric medium. Its use --- indeed, even its validity --- in a science increasingly embracing the Principle of Relativity was therefore considered questionable. It attracted considerably less interest than Abraham's form, which was developed in a fully relativistic manner, working from microscopic considerations.

In contrast,
\textcite{dallenbach} claimed to demonstrate the derivation of the Minkowski tensor from microscopic considerations.
However, according to
\textcite{degroot} he fails to justify his generalisation from the electrostatic to the dynamic case,
and
\textcite[p.109]{pauli2}
also deemed his argument ``not very cogent''.

As we noted in Sec.~\ref{earlyyears}, concerns about non-conservation of angular momentum in the Minkowski electromagnetic energy--momentum tensor arose from consideration of an incomplete system. When the appropriate material tensor is also taken into account, angular momentum is conserved (Sec.~\ref{proofsequiv}). Similarly the Einstein--Laub tensor can be saved from the ignonimy of depending upon a specific frame of reference.

Later, \textcite{vonlaue}
argued that the velocity of energy propagation, which is less than $c$ in a dielectric medium, should add in accordance with the relativistic velocity addition theorem.
An equivalent argument,
developed by \textcite[pp. 206-211 and 221-225 respectively]{moller,moller2}, is that the energy propagation velocity should transform as a 4-vector.
This is satisfied by the electromagnetic energy flux under the Minkowski tensor, but not under the Abraham tensor. In the first edition of his work M\o{}ller concludes  ``\ldots{} this is a strong argument in favour of Minkowski's theory'', but in the second edition this is softened to ``\ldots{} which shows that Minkowski's theory gives the most convenient description of the phenomenon in question''. This alteration perhaps reflects a change in philosophy, 
away from attempting to prove that one tensor is wrong and the other is right, and towards recognising that both have their place in a practical implementation of electrodynamics.

Although it was ultimately dismissed by \textcite{tang} and \textcite{skobeltsyn},
von Laue's argument was nevertheless influential.
When Pauli first wrote his encyclopedia article on relativity \cite{pauli},
he favoured the Abraham tensor. However, in the
revised edition \cite{pauli2}, he inserted a supplementary note
reversing his position on account of von Laue's argument.

In addition to rebutting the arguments of von Laue and M\o{}ller, Skobel'tsyn also proposed criteria of his own based on two thought experiments, which favoured the Abraham tensor. 
These are described in detail by \textcite{brevik3}, along with his reasons for rejecting Skobel'tsyn's arguments. Brevik then concludes that the second thought experiment, which deals with a capacitor free to rotate within a magnetic field, would nevertheless provide an experimental means to discriminate between the Abraham and Minkowski tensors.
The behaviour of the experiment depends upon the existence or otherwise of the Abraham force, as does the experiment of \textcite{walker}. The rebuttals of \textcite{israel} and \textcite{obukhov} discussed in Sec.~\ref{expts:walker} are equally applicable to Skobel'tsyn's second experiment, and hence both Skobel'tsyn's arguments and Brevik's interpretation of the second experiment are refuted.

\textcite{lai}
also published an argument against the Minkowski tensor based on non-conservation of angular momentum, similar to Skobel'tsyn's second argument. This was rebutted by
\textcite{brevik4} by invoking the material counterpart to the Minkowski electromagnetic tensor. We raise this here to discuss Lai's counter-argument \cite{lai2}. In this, he argues that his thought experiment was not intended to distinguish between what he calls the ``phenomenological'' theories derived from global conservation of momentum, but rather to determine which of Abraham's and Minkowski's tensors is ``physically correct''. How this is to be interpreted is unclear.

If Lai is advocating comparison of the historical forms of these tensors, then as neither incorporates effects such as electrostriction and magnetostriction, both are incomplete and demonstrably incorrect. 
We can only ask which is least inaccurate in any particular situation, a conclusion of little general value.

If Lai permits the material tensor accompanying the Abraham tensor to be amended to incorporate such effects, but does not permit the Minkowski tensor to receive the same treatment, then he is crippling one of the contenders before 
the competition,
and to conclude that one then performs better than the other is hardly surprising.

Finally, if both tensors are allowed such a counterpart, then we recover the treatment of Brevik, which Lai dismisses as ``phenomenological'' and missing the point.

Alternatively, Lai may be implying that it is possible to discriminate between the Minkowski and Abraham tensors when both are accompanied by the appropriate material tensors. This position, however, has been disproven by others \cite{mikura,kranys,kranys3,maugin,kranys2,schwarz}, and
hence Lai's thought experiment appears to serve no useful purpose, in spite of his defence.

It is of historical note that the Abraham tensor was also initially proposed without a material counterpart \cite{abraham,abraham2,brevik2}, and it was several decades before the role of this counterpart in determining experimental results was properly appreciated (see Sec.~\ref{expts:ashkin}).
Any treatment involving such a material counterpart is to some extent phenomenological, with the nature of the counterpart being determined or verified by observable fact \cite{jones,jones2}.
It can be argued that only approaches which begin from universal conservation laws may be exempt from this criticism, if criticism it is, but it is these very results which Lai appears to dismiss out of hand.

Other arguments purportedly favouring one tensor or the other had also been proposed
\cite{gyorgyi,tolman,tamm}, but ultimately it was not to be these which decided the debate. In the 1960s and 1970s, two developments occurred which changed the face of the argument completely.
First, a number of further alternative energy--momentum tensors were proposed on various theoretical grounds \cite{penfield,grot,degroot,degroot2a,degroot4,marx,peierls}.
The debate was no longer over whether to choose Abraham's or Minkowski's formulation, but rather over the principles upon which a valid electromagnetic energy--momentum tensor should be based. Second, 
\textcite{blount} identified the Abraham electromagnetic energy--momentum tensor with classical momentum and the Minkowski tensor with crystal momentum, or pseudomomentum. Although the work was unpublished it had considerable effect, stimulating a recognition that both energy--momentum tensors might have a valid role in appropriate circumstances.

Although not every proposed means of discriminating between the Abraham and Minkowski tensors has been explicitly refuted in the literature, there is little to be gained from such a process. Inevitably, such attempts stem from inadequate exploration of the material energy--momentum tensor, and ultimately all such arguments are overturned by the direct proofs of equivalence of the differing electromagnetic energy--momentum tensors discussed in Sec.~\ref{proofsequiv}. These proofs arise directly from conservation of energy, linear and angular momentum for the system as a whole, and hence it is only in those rare papers which discard these constraints that differing predictions can be made in a logically consistent manner.


\section{Proofs of Equivalence\label{proofsequiv}}

\subsection{Penfield and Haus; de Groot and Suttorp\label{proofs1}}

Towards the end of the 1960s, a number of researchers were demonstrating greater awareness of the role of the material energy--momentum tensor in symmetry and in momentum conservation \cite{degroot2a,grot}, and developing novel pairs of electromagnetic and material tensors with this in mind.
\textcite{penfield}
took this further with the formulation of their ``Principle of Virtual Power'', developed as a tool to test the equivalence of a number of different formulations of electromagnetic theory. By applying this principle they were able to determine the incompleteness of both the Minkowski and Abraham tensors, and also of the tensors emerging from the 
\textcite{boffi}, Amperian \cite{fano} and Chu \cite{penfield,chu}
formulations of electromagnetism, and to identify the missing terms, and demonstrate that with properly constructed material counterparts, all these energy--momentum tensors are in agreement. A cogent summary of their arguments and results may be found in a 
review paper by
\textcite{robinson}, which also covers the approach of Gordon, discussed in Sec.~\ref{sec:gordon}.

Like Penfield and Haus,
\textcite{degroot}
also recognised the important role of the material energy--momentum tensor, and that conservation laws relating to energy and momentum should only be applied to closed thermodynamic systems. However, they disagree with Penfield and Haus on the correct form of the \emph{total} energy--momentum tensor to which electromagnetic and material energy--momentum tensors must sum. This disagreement does not bear directly upon the Abraham--Minkowski controversy, but does reflect different overall assumptions about the behaviour of matter in the presence of an electromagnetic wave. De Groot and Suttorp have argued for the superiority of their expression as it is derived from considerations of the microscopic properties of the medium. As yet, no experimental comparison of these two different total energy--momentum tensors has been performed.

For the moment, however, the correct expression for the total energy--momentum tensor is unimportant. While eventually it might prove beneficial to our modelling of material behaviour to discriminate between the different propositions, for our purposes it suffices to denote the selected total energy--momentum tensor by $T^{\alpha\beta}$. This tensor is then constrained by conservation of linear and angular momentum such that
\begin{equation}
\partial_\alpha T^{\alpha\beta}=0\label{nodiv}
\end{equation}
and
\begin{equation}
T^{\alpha\beta}=T^{\beta\alpha},\label{symmetry}
\end{equation}
\cite{jackson}, where greek indices range over 0-3.

A more familiar expression of the conservation of linear momentum may be
\begin{equation}
\partial_t \, \mbf{g}^{i} + \partial_{j} \, \mbf{T}^{{ij}} = 0,\label{linmomcons}
\end{equation}
where latin indices range over 1-3, $\partial_t$ represents the derivative with respect to time, \textbf{g} is the momentum density, and \textbf{T} is the three-dimensional stress tensor. The stress tensor $\mbf{T}$ is related to the energy--momentum tensor $T$ via
\begin{equation}
\mbf{T}^{ij}=-T^{ij}.
\end{equation}
For an electromagnetic wave in free space, \textbf{T} is called the Maxwell stress tensor, but in dielectric media two different conventions exist \cite{stallinga2} so unless explicitly defined, this nomenclature is best avoided.

A further constraint
\begin{equation}
\partial_t \, g^0 - \partial_{j} \, T^{0j} = 0
\end{equation}
corresponds to the conservation of energy,
where $g^0$ is the energy density, and $T^{0j}$ is the energy flux. For an electromagnetic wave in free space, $T^{0j}=\mbf{S}^j/c$ where $\mbf{S}$ is the Poynting vector. Identifying $(g^0,\mbf{g})$ as a 4-vector and writing
\begin{equation}
T^{\alpha\beta} = \left( \begin{array}{c|c}
g^0 & T^{0j} \\
\hline
c\mbf{g}^{i} & -\mbf{T}^{ij}
\end{array}\right),
\end{equation}
we see that (\ref{linmomcons}) is equivalent to constraint (\ref{nodiv}) above.
Defining $\mbf{g}^\prime = ({1}/{c})\,T^{0j}$ for symmetry, constraint (\ref{symmetry}) requires that $\mbf{g}=\mbf{g^\prime}$, giving
\begin{equation}\label{findgT}
T^{\alpha\beta} = \left( \begin{array}{c|c}
g^0 & c\mbf{g}^{j} \\
\hline
c\mbf{g}^{i} & -\mbf{T}^{ij}
\end{array}\right).\label{eq:Tdivide}
\end{equation}

Unlike the system described by the total energy--momentum tensor, the one described by the electromagnetic energy--momentum tensor need not be thermodynamically closed. Therefore, one must consider the combined system of both the electromagnetic and material energy--momentum tensors, such that
\begin{equation}
T^{\alpha\beta}=T^{\alpha\beta}_{\mathrm{EM}}+T^{\alpha\beta}_{\mathrm{mat}} \label{construct}
\end{equation}
and write
\begin{eqnarray}
\partial_\alpha \left( T^{\alpha\beta}_{\mathrm{EM}} + T^{\alpha\beta}_{\mathrm{mat}}\right)&=&0 \label{eq:nodiv}\\
\left( T^{\alpha\beta}_{\mathrm{EM}} + T^{\alpha\beta}_{\mathrm{mat}}\right) &=& \left( T^{\beta\alpha}_{\mathrm{EM}} + T^{\beta\alpha}_{\mathrm{mat}}\right). \label{symmetry2}
\end{eqnarray}
As before, we may rewrite (\ref{construct}) and (\ref{eq:nodiv}) in terms of \textbf{g} and \textbf{T}:
\begin{eqnarray}
\mbf{g}&=&\mbf{g}_\mrm{EM} + \mbf{g}_\mrm{mat}\label{eq:g-subdiv}\\
\mbf{T}&=&\mbf{T}_\mrm{EM} + \mbf{T}_\mrm{mat}
\end{eqnarray}
\begin{eqnarray}
\partial_t \, (\mbf{g}^{i}_\mrm{EM} + \mbf{g}^{i}_\mrm{mat}) + \partial_{j} \, (\mbf{T}^{ij}_\mrm{EM} + \mbf{T}^{ij}_\mrm{mat}) &=& 0\\
\partial_t \, ({g^0}_\mrm{EM} + {g^0}_\mrm{mat}) - \partial_{j} \, ({T}^{0j}_\mrm{EM} + {T}^{0j}_\mrm{mat}) &=& 0.
\end{eqnarray}
Again we can define
$\mbf{g}^\prime = (1/c)T^{0j},$
giving
\begin{eqnarray}
T &=& T_\mrm{EM} + T_\mrm{mat}\\
&=&\left( \begin{array}{c|c}
g^0_\mrm{EM}+g^0_\mrm{mat} & c{\gprime}^{j}\\
\hline
c(\mbf{g}_\mrm{EM}^{i} + \mbf{g}_\mrm{mat}^{i}) & -(\mbf{T}_\mrm{EM}^{ij} + \mbf{T}_\mrm{mat}^{ij})
\end{array}\right).
\end{eqnarray}
$\gprime$ must also be subdivided,
\begin{equation}
\gprime=\gprime_{\mrm{EM}}+\gprime_\mrm{mat},
\end{equation}
and from (\ref{symmetry2}) we see that
\begin{equation}
\gprime_{\mrm{EM}}+\gprime_\mrm{mat} = \mbf{g}_\mrm{EM} + \mbf{g}_\mrm{mat},
\end{equation}
but no constraint is placed upon the subdivision of $\mbf{g}^\prime$ into electromagnetic and material components. In energy--momentum tensors where the electromagnetic portion is symmetric, such as the Abraham tensor, it follows that
\begin{eqnarray}
\mbf{g}_\mrm{EM}&=&\mbf{g}^\prime_\mrm{EM}\\
\mbf{g}_\mrm{mat}&=&\mbf{g}^\prime_\mrm{mat},
\end{eqnarray}
but this is not obligatory. For the Minkowski tensor, $\mbf{g}_\mrm{EM}=\mbf{D}\times\mbf{B}$, whereas $\mbf{g}^\prime_\mrm{EM}=(1/c^2)\mbf{E}\times\mbf{H}$. Similarly, $\mbf{g}_\mrm{mat}\not=\gprime_\mrm{mat}$.

When the Minkowski tensor is used, $T_\mrm{mat}$ is frequently ignored as the material components of $\mbf{g}$ and $\mbf{T}$ are typically safely negligible (see Sec.~\ref{discussion}).
However, because $\mbf{g}_\mrm{EM}\not=\gprime_\mrm{EM}$ under the Minkowski tensor, this gives the illusion of non-conservation of angular momentum via (\ref{symmetry}). The Minkowski electromagnetic energy--momentum tensor has received much criticism for this reason, due to failure to recognise the existence of its material counterpart.

Historically, neither the Minkowski nor the Abraham tensors were proposed in the context of a matched pair of electromagnetic and material energy--momentum tensors, and although the need for a material counterpart to the Abraham tensor was proposed by \textcite{jones}, the Minkowski tensor only acquired a material counterpart through the theoretical re-evaluations of the 1970s \cite{degroot,mikura,israel}, explaining the results of \textcite{james} and \textcite{walker} discussed in Sec.~\ref{expts:walker}.


We can now explain why experiments continue to be announced from time to time which purport to discriminate between the Minkowski and Abraham energy--momentum tensors, only to be disproven a few years later. For brevity we adopt the notation of Eqs. (\ref{nodiv}) and (\ref{symmetry}). Any experiment will be sensitive only to certain terms in the complete energy--momentum tensor. An experiment involving only angular momentum will be insensitive to independently symmetric terms, and one involving only linear momentum will be insensitive to terms which are independently divergence-free. For any given experiment, we may therefore define $T_\mathrm{X}$ as the terms in the material tensor which have no significant effect on the result, and $T_\mathrm{mat^{\prime}}$ as
\begin{equation}
T_\mathrm{mat^{\prime}}=T_\mathrm{mat}-T_\mathrm{X}.
\end{equation}
Using $T_\mathrm{mat^{\prime}}$ in lieu of $T_\mathrm{mat}$ to make predictions for the behaviour of one experiment will therefore still yield the correct result, but for another experiment it may not. To illustrate this, consider the Minkowski electromagnetic energy--momentum tensor. In general, this tensor is used in isolation, letting $T_\mathrm{mat^{\prime}}=0$. The Minkowski tensor is asymmetric, and hence from Eqs.~(\ref{symmetry}) and (\ref{symmetry2}) we can conclude that $T_\mathrm{X}$ also contains asymmetric terms. The Minkowski electromagnetic energy--momentum tensor largely functions well in linear experiments such as that of Jones and Richards (Sec.~\ref{expts:jones} above), but in isolation fails to 
correctly describe the rotary experiment of 
Walker \emph{et al.} (Sec.~\ref{expts:walker}).
\textcite{israel} and \textcite{obukhov}
demonstrated that this is due to neglect of the corresponding material tensor. 

A given electromagnetic and material tensor pair may be found to perform well across a broad range of experiments before failing. This failure can arise as a result of an interaction which was not probed in the previous experiments, and is therefore described by a term which does not influence those previous experiments. A frequent reaction is to announce that the experiment invalidates one of the electromagnetic energy--momentum tensors being considered. However, inevitably a suitable term will be identified and added on to the corresponding material energy--momentum tensor, or a new expression developed which is equivalent to the addition of such a term, and the new ``correct'' material energy--momentum tensor will be announced, rebutting the invalidation
\cite{brevik4,israel,obukhov}.
As the tensor pair is named by the structure of the electromagnetic member, and the necessary term may always be added to the material member, the tensor is therefore always able to be ``saved''.

As an interesting alternative, the necessary term may be added to the electromagnetic tensor, and accompanied by a change of name. Neither the Abraham nor Minkowski tensors as they are commonly used incorporate the electrostrictive effect. When the relevant term is added to the electromagnetic component of the Abraham tensor, this is known as the Helmholtz tensor \cite{brevik4}.

Regrettably, the full ramifications of the works of Penfield, Haus, de Groot and Suttorp were not immediately appreciated by the physics community, perhaps because calculating the full expressions for the material energy--momentum tensors was somewhat laborious, and perhaps because the experiment of
\textcite{walker}
had yet to be performed, that of
\textcite{james,james2}
was not widely publicised, and thus no convincing situation had yet been recognised for which the conventional forms of the Minkowski and Abraham tensors failed or disagreed.

In addition, de Groot and Suttorp derived a number of new tensor pairs which they felt were particularly relevant to specific situations, such as a neutral plasma, and which are documented in their book \cite{degroot}.

\subsection{Gordon's Analysis\label{sec:gordon}}

A more practically oriented demonstration of the equivalence of the Abraham and Minkowski tensors was achieved by
\textcite{gordon}. In response to the experiment of Ashkin and Dziedzic (see Sec.~\ref{expts:ashkin} above), he demonstrated that 
the field-related components of the material energy--momentum tensor
propagated along with the electromagnetic component, rather than at the speed of sound in the medium as was sometimes assumed
\cite[for example][]{burt},
and that the optical pulse length in Ashkin and Dziedzic's experiment was sufficiently long for pressure within the fluid to reach equilibrium, and so purely material effects such as pressure also needed to be considered. 
He then went on to show 
that both Abraham's and Minkowski's tensors predicted identical behaviour when an arbitrary radiation pulse traversed an air/fluid interface.

We note that all real material bodies are to some extent capable of deformation, and hence allow the transmission of tension and pressure in a manner
analogous to the effects considered in Gordon's proof. The proof therefore 
readily generalises to arbitrary isotropic dielectric media for deformations small enough that dissipative effects such as hysteresis may be ignored,
and has the potential for extrapolation to larger deformations and anisotropic cases. Inhomogeneous dielectric materials might then be dealt with by breaking them down into infinitesimally small homogeneous chunks.
However, as pointed out by
\textcite{robinson},
Gordon's result is only explicitly demonstrated for materials of dielectric constant sufficiently close to unity.

Gordon's work was subsequently extended by
\textcite{peierls,peierls2}
with a number of interesting consequences, such as the generation of phonons by a finite width beam in an elastic solid.

Recognition of the deformable nature of the dielectric medium is fundamental to Gordon's approach.
We discuss the hazards associated with inappropriately treating the dielectric as a rigid body
in Sec.~\ref{anomeffects}.

\subsection{Mikura, Krany\v{s} and Maugin}

Although the work of
\textcite{degroot}
had provided the theoretical framework within which to demonstrate the formal equivalence of different choices of 
electromagnetic energy--momentum tensor, they stopped short of explicitly demonstrating this for commonly used tensors such as those of Abraham and Minkowski. This equivalence had been demonstrated by
Penfield and Haus in 1967,
but it seems to have been little appreciated.

Subsequent, far shorter demonstrations of equivalence were produced by
\textcite{mikura}
and 
by
\textcite{kranys,kranys2,kranys3},
who appears unaware of the work of either Penfield and Haus or de Groot and Suttorp, as he cites later papers by
\textcite{israel,israel2}
as conjecturing a ``sliding'' relationship between electromagnetic and material energy--momentum tensors. The paper by Mikura is particularly readable, but both are recommended to
anyone seeking greater mathematical detail than is provided in this review.
\textcite{ginzburg}
also demonstrates that the behaviours predicted by the classical Abraham and Minkowski tensors are equivalent up to the action of the Abraham force, but does not take the final step of postulating a material counterpart to the Minkowski tensor. He therefore concludes that while the Minkowski tensor is not strictly correct, it is nevertheless useable under many circumstances.

A response to \textcite{kranys} by
\textcite{maugin} supports Krany\v{s}'s result but criticises him for not including enough references, and for confining himself to considering only the
Abraham and Minkowski tensors. Maugin extends Krany\v{s}'s work to consider tensors due to
\textcite{grot} and \textcite{degroot}
in non-dissipative materials, with remarks about extension to more arbitrary cases, and the use of this approach to demonstrate the equivalence of any other energy--momentum tensor with those already considered.
\textcite{schwarz} then
extends the work of Krany\v{s} and Maugin to include media which may be magnetised and/or electrically polarised.

The interested reader may also have inferred the general 
equivalence of all choices of electromagnetic energy--momentum tensor from 
Eqs.~(\ref{nodiv}), (\ref{symmetry}) and (\ref{construct}),
as valid expressions for the total energy--momentum tensor $T$ may only vary by a term which is symmetric and divergence free, and hence carries no momentum at all.

\section{Momentum and Pseudomomentum\label{mompseudomom}}

As mentioned in Sec.~\ref{arguments}, an unpublished memorandum by \textcite{blount} is credited with first drawing the parallel between the Minkowski electromagnetic momentum density and pseudomomentum. In doing so, he introduced a viewpoint in which both the Abraham and Minkowski expressions had a valid role. This view is further developed by \textcite{gordon}, \textcite{peierls3},
\textcite{nelson}, and \textcite{stallinga},
and our review would be incomplete without discussing this important recognition.

\textcite{balazs} provides a thought experiment which we may use 
to illustrate this. Consider a perfectly transparent glass rod, initially at rest, with a perfect anti-reflection coating on each end. A wave packet may pass either alongside the rod, or through it. No external force acts on the combined system of rod and wave packet, so the centre of mass of the system must have constant velocity. However, if the wave packet enters the rod, it slows down by a factor of $1/n$. This must therefore be accompanied by a movement of the rod in the direction of propagation of the wave packet so that the velocity of the centre of mass is preserved. As the rod is perfectly transparent, no momentum is absorbed from the wave packet, and when the wave packet departs the rod it must leave the rod at rest. By comparing the trajectories of the centre of mass with the wave packet passing either inside or outside the glass rod, we can show that the rod experiences a finite displacement during the traversal of the wave packet, which we may interpret as resulting from the action of the Lorentz force on the atomic dipoles which make up the rod.

While the rod is moving, it clearly has momentum. The Abraham approach identifies this as mechanical momentum. However, the Minkowski approach recognises that this momentum propagates with and arises from the presence of the electromagnetic wave, and hence classifies it as electromagnetic momentum.

We may well have qualms about describing the motion of a material body as electromagnetic momentum. Therefore, we may call the Abraham momentum the true momentum of an electromagnetic wave, and the Minkowski momentum the pseudomomentum, this being the physically useful quantity of all momentum associated with the propagation of the wave. (Nelson adopts a different terminology, in which pseudomomentum excludes electromagnetic momentum, and the total momentum associated with the propagation of a wave is termed ``wave momentum''. His chosen expression for electromagnetic momentum also differs, though this has no effect on the total momentum. Nelson identifies the Minkowski momentum with wave momentum, less a term due to dispersion.)


This argument may appear compelling, but before the reader confidently declares the Abraham momentum to be the only true electromagnetic momentum, they should recall that every expression appears compelling within the context of the arguments on which it is founded. 
For example, \textcite{mansuripur} presented an alternative chain of argument leading to an electromagnetic momentum density of $\mbf{g}=\frac{1}{2}\mbf{D}\times\mbf{B} + \frac{1}{2}\mbf{E}\times\mbf{H}/c^2$. This expression seems unlikely in the context of the above discussion of momentum and pseudomomentum, but in the approach taken by Mansuripur it is the Abraham and Minkowski expressions which appear inappropriate, and this hybrid electromagnetic momentum density which seems the inevitable conclusion.
Of course, it is momentum transfer which determines experimental results, and this takes place identically regardless of whether we describe it as electromagnetic or material in nature. We therefore reprise our position stated in the abstract, that the reader may describe as electromagnetic whatever portion of the total momentum flux appears most appropriate, in the context in which they are working. Provided all momentum is ultimately accounted for, 
this will have no effect on their predicted experimental results.

\textcite{gordon} has given the conditions under which the pseudomomentum approach will reliably describe momentum transfer to an object in a fluid dielectric medium. The factor $|\mbf{E}|^2\nabla\varepsilon$ must be negligible in the vicinity of a surface surrounding the object of interest, and sufficient time must be allowed for pressure in the fluid to reach equilibrium. In contrast the total energy--momentum tensor approach described in Sec.~\ref{totememt} may be employed even when the constituent variables are largely arbitrary functions of space and time, though the cost is a considerable increase in mathematical complexity.

Performing calculations using the pseudomomentum density $\mbf{D}\times\mbf{B}$ (in Gordon's notation and Gaussian units, $\epsilon\mbf{G}$) is directly equivalent to choosing the Minkowski electromagnetic energy--momentum tensor without a material counterpart. 
It therefore omits a small number of effects associated with the material counterpart tensor, including the Abraham force, but these are seldom significant. Situations 
suited to the use of the isolated Minkowski electromagnetic tensor, and hence also the pseudomomentum approach, are discussed in Sec.~\ref{ex:mirror} and \ref{discussion}.

\section{The Total Energy--Momentum Tensor\label{totememt}}

The obvious question now becomes, what is the total energy--momentum tensor? Can we write down a single expression that will serve our needs in all circumstances? Unfortunately, the answer is no. This is not a problem in electromagnetism. The limitation is instead our understanding of material science. We may certainly write down all contributions to the energy--momentum tensor which are relevant in a particular perfectly defined ideal medium, but to do so for a generalised real medium 
lies beyond the scope of our current understanding.

Fortunately, in many circumstances a complete understanding of the material properties of the medium is not required. The medium may behave sufficiently like a suitable ideal, or we may be interested solely in the behaviour of the centre of mass, and we may then neglect the effects of internal tensions which merely redistribute existing force and torque densities. Provided our description of the interaction of the material object with the electromagnetic wave is sufficiently accurate, details of material-material interactions may then be overlooked.

There is another important caveat.
When material stresses induced by an electromagnetic wave cause deformation of the boundary of the dielectric medium, focussing effects may then alter the subsequent path of the beam, as in the experiment of \textcite{ashkin}. 
The results calculated for a given material conformation will then only be instantaneously true. This effect may of course neglected if the deformation of the material medium is sufficiently small.

\subsection{A Total Energy--Momentum Tensor\label{sec:atotal}}

\textcite{mikura} derives a total energy--momentum tensor for a nonviscous, compressible, nondispersive, polarisable, magnetisable, isotropic fluid. Electrostrictive effects, magnetostrictive effects, and acoustic waves are included, though in most circumstances they may be omitted for an increase in simplicity, as shown by Mikura in the latter part of his paper. 
We have converted from Mikura's choice of Gaussian units and matrix notation to SI units and tensor notation. Mikura's approach may also be extended to dispersive media, as discussed in Sec.~\ref{qualdis}. We will therefore distinguish between the phase and group refractive indices, denoted $n_\phi$ and $n_g$ respectively.

The total energy--momentum tensor is given by
\begin{equation}
T^{\mu\nu}=T^{\mu\nu}_{(m)}+T^{\mu\nu}_{(f)}+T^{\mu\nu}_{(P)}+T^{\mu\nu}_{(M)}+T^{\mu\nu}_{(d)}
\end{equation}
where
\begin{eqnarray}
T^{\mu\nu}_{(m)}&=&\left( \rho_0 c^2+\rho_0 \epsilon_\mrm{i}\right) u^\mu u^\nu + \phi \left( u^\mu u^\nu +\delta^{\mu\nu} \right)\nonumber\\
T^{\mu\nu}_{(f)}&=&\varepsilon_0 c^2 \left( F^{\mu\gamma} {F^\nu}_\gamma - \frac{1}{4} F^2_{\gamma\delta} \,\delta^{\mu\nu}\right)\nonumber\\
T^{\mu\nu}_{(P)}&=&\frac{1}{\alpha}\left(P^{\mu\gamma} {P^\nu}_\gamma - \frac{1}{4} P^2_{\gamma\delta} \,\delta^{\mu\nu}\right)\nonumber\\
T^{\mu\nu}_{(M)}&=&-F^{\mu\gamma} {M^{*\,\nu}}_\gamma - \frac{1}{4\beta} M^2_{\gamma\delta} \,\delta^{\mu\nu}\nonumber\\
T^{\mu\nu}_{(d)}&=& \frac{1}{\beta}M^{\mu\gamma} {M^\nu}_\gamma + F^{*\,\mu\gamma} {M^\nu}_\gamma.\nonumber
\end{eqnarray}
We have slightly condensed Mikura's original scheme. $T_{(m)}$ contains terms due to mechanical fluid flows, electrostriction, and magnetostriction. $T_{(f)}$ contains field terms identical to those in free space. $T_{(P)}$ contains terms relating to the polarisation of the medium, and $T_{(M)}$ and $T_{(d)}$ contain terms relating to magnetisation. $T_{(d)}$ is separated from $T_{(M)}$ to facilitate the subsequent derivation of the Abraham and Minkowski electromagnetic energy--momentum tensors.

In these expressions, $\rho_0$ is the matter density in the local rest frame, $u$ is the 4-velocity of an element of the local medium, $\epsilon_\mrm{i}$ is the specific internal energy of non-electromagnetic nature and is a function of $\rho_0$ and the specific elastic entropy $s$, $\alpha$ and $\beta$ are related to the electric and magnetic constants of the medium by $\alpha=\varepsilon-\varepsilon_0$ and $\beta=\mu_0^{-1}-\mu^{-1}$, $\delta^{\mu\nu}=1$ for $\mu=\nu$ and zero otherwise, 
$\phi$ is the total pressure including electrostrictive and magnetostrictive effects, given by
\begin{equation}
\phi=\phi_\mrm{h}-\frac{1}{4}K_\mrm{a} P^2_{\gamma\delta} + \frac{1}{4}K_\mrm{b} M^2_{\gamma\delta}
\end{equation}
for
\begin{equation}
K_\mrm{a}=\rho_0\left.\frac{\partial(1/\alpha)}{\partial\rho_0}\right|_s,\quad K_\mrm{b}=\rho_0\left.\frac{\partial(1/\beta)}{\partial\rho_0}\right|_s,\\
\end{equation}
and $\phi_\mrm{h}$ is the total hydrostatic pressure in the fluid,
\begin{equation}
\phi_\mrm{h}=\rho_0^2\left.\frac{\partial\epsilon_\mrm{i}}{\partial\rho_0}\right. .
\end{equation}
$F$ is the usual electromagnetic field tensor
\begin{samepage}
\begin{equation}
F^{\mu\nu}=\left( \begin{array}{cccc}
0&-\frac{E_1}{c}&-\frac{E_2}{c}&-\frac{E_3}{c}\\
\frac{E_1}{c}&0&-B_3&B_2\\
\frac{E_2}{c}&B_3&0&-B_1\\
\frac{E_3}{c}&-B_2&B_1&0
\end{array}\right),
\end{equation}
and the polarisation and magnetisation tensors $P$ and $M$ are related to the three-dimensional polarisation and magnetisation vectors $\mbf{P}$ and $\mbf{M}$ in a given frame by
\end{samepage}
\begin{eqnarray}
P^{\mu\nu}&=&(1/c)(\mbf{v}^\mu\mbf{P}^\nu-\mbf{v}^\nu\mbf{P}^\mu)\\
M^{\mu\nu}&=&(1/c)(\mbf{v}^\mu\mbf{M}^\nu-\mbf{v}^\nu\mbf{M}^\mu)
\end{eqnarray}
where $\mbf{v}$ is the 3-velocity of an element of the local medium. 
We define
$\mbf{v}^0=c$, $\mbf{P}^0=\mbf{M}^0=0$. $\mbf{P}$ and $\mbf{M}$ are related in turn to $\mbf{E}$, $\mbf{B}$, $\mbf{D}$, and $\mbf{H}$ via $\mbf{D}=\varepsilon\mbf{E}$, $\mbf{B}=\mu\mbf{H}$, and the relativistic constitutive relations
\begin{eqnarray}
\mbf{D}
&=&\varepsilon_0\mbf{E}+\mbf{P}+\frac{1}{c}\left(\frac{\mbf{v}}{c}\times\mbf{M}\right)\\
\mbf{H}
&=&\frac{1}{\mu_0}\,\mbf{B}-\mbf{M}+c\left(\frac{\mbf{v}}{c}\times\mbf{P}\right).
\end{eqnarray}

For \textit{F}, \textit{M}, and \textit{P}, the notation $X^2_{\gamma\delta}$ abbreviates $X^{\gamma\delta}X_{\gamma\delta}$, and $X^*$ denotes the dual tensor
\begin{equation}
X^{*\,\mu\nu} = \frac{1}{2}\,\varepsilon^{\mu\nu\gamma\delta} X_{\gamma\delta}
\end{equation}
where $\varepsilon$ is the Levi--Civit\`{a} symbol.

The total momentum flux and stress tensor under this total energy--momentum tensor may be obtained via $\mbf{g}^i=\frac{1}{c}T^{i0}$ and $\mbf{T}^{ij}=-T^{ij}$.
\begin{widetext}
\begin{eqnarray}
\mbf{g}&=&[\rho_0 (c^2+\epsilon_\mrm{i})+\phi ]\gamma^2\frac{\mbf{v}}{c}
+ \varepsilon_0\,\mbf{E}\times\mbf{B} 
+ \frac{1}{c\alpha}\,\mbf{P}\times\left(\frac{\mbf{v}}{c}\times\mbf{P}\right)
+ \frac{1}{c\beta}\,\mbf{M}\times\left(\frac{\mbf{v}}{c}\times\mbf{M}\right)
\nonumber\\
&& - \frac{1}{c}\,\mbf{B}\times\left(\frac{\mbf{v}}{c}\times\mbf{M}\right)
+\frac{1}{c^2}\,\mbf{E}\times\mbf{M}\\
\mbf{T}&=&-\rho_0(c^2+\epsilon_\mrm{i})\gamma^2\frac{\mbf{v}\wedge \mbf{v}}{c^2}
- \phi\left(\gamma^2\frac{\mbf{v}\wedge\mbf{v}}{c^2}+\mbf{I}\right)
%
 + \varepsilon_0\mbf{E}\wedge\mbf{E} 
  + \frac{1}{\mu_0}\,\mbf{B}\wedge\mbf{B} 
  - \frac{1}{2}\left(\varepsilon_0|\mbf{E}|^2+\frac{1}{\mu_0}\,|\mbf{B}|^2\right)\mbf{I}\nonumber\\
%
%
&& + \frac{1}{\alpha} \left[ \mbf{P}\wedge\mbf{P} 
 + \left( \frac{\mbf{v}}{c}\times\mbf{P} \right)\wedge \left( \frac{\mbf{v}}{c}\times\mbf{P} \right)
 -\frac{1}{2}\left(\left|\frac{\mbf{v}}{c}\times\mbf{P}\right|^2
 - |\mbf{P}|^2\right)\mbf{I}\right]\nonumber\\
&& + \frac{1}{\beta}\left[\mbf{M}\wedge\mbf{M}
 + \left(\frac{\mbf{v}}{c}\times\mbf{M}\right)\wedge \left(\frac{\mbf{v}}{c}\times\mbf{M}\right)
 - \frac{1}{2}\left(\left|\frac{\mbf{v}}{c}\times\mbf{M}\right|^2
 - |\mbf{M}|^2\right)\mbf{I}\right]\\
&& - \mbf{B}\wedge\mbf{M} - \mbf{M}\wedge\mbf{B}
+ \frac{\mbf{E}}{c} \wedge \left(\frac{\mbf{v}}{c}\times\mbf{M}\right)
+ \left(\frac{\mbf{v}}{c}\times\mbf{M}\right) \wedge \frac{\mbf{E}}{c}
- \left[\frac{\mbf{E}}{c}\cdot\left(\frac{\mbf{v}}{c}\times\mbf{M}\right)
 - \mbf{B}\cdot\mbf{M}\right]\mbf{I}.\nonumber
\end{eqnarray}
\end{widetext}
$\mbf{x}\wedge\mbf{y}$ denotes the outer product, $(\mbf{x}\wedge\mbf{y})^{ij}=x^iy^j$.

Note that the trace over all field-related components of the total energy--momentum tensor vanishes, and it is therefore compatible with massless photons \cite{jackson}.


Following Mikura, we also subdivide this tensor to yield the Abraham and Minkowski electromagnetic and material energy--momentum tensor pairs. 
For the Abraham tensor, we simply write
\begin{eqnarray}
T^{\mu\nu}_{\mrm{mat, Abr}}&=&T^{\mu\nu}_{(m)}\\
T^{\mu\nu}_{\mrm{EM, Abr}}&=&T^{\mu\nu}_{(f)}+T^{\mu\nu}_{(P)}+T^{\mu\nu}_{(M)}+T^{\mu\nu}_{(d)}.
\end{eqnarray}
To obtain the Minkowski tensor we 
define
\begin{equation}
C^{\mu\nu}=\frac{1}{\alpha}P^{\mu\gamma}{P^\nu}_\gamma - F^{\mu\gamma}{P^\nu}_\gamma
\end{equation}
and then we find
\begin{eqnarray}
T^{\mu\nu}_{\mrm{mat, Mink}}&=&T^{\mu\nu}_{(m)}+T^{\mu\nu}_{(d)}+C^{\mu\nu}\\
T^{\mu\nu}_{\mrm{EM, Mink}}&=&T^{\mu\nu}_{(f)}+T^{\mu\nu}_{(P)}+T^{\mu\nu}_{(M)}-C^{\mu\nu}.
\end{eqnarray}
In the nonrelativistic limits, these yield the 
usual expressions for the electromagnetic energy--momentum tensors:
\begin{widetext}
\begin{eqnarray}
T^{\mu\nu}_{\mrm{EM, Abr}}&=&\left( \begin{array}{cc}
\frac{1}{2}\left(\mbf{E}\cdot\mbf{D}+\mbf{H}\cdot\mbf{B}\right) & \frac{1}{c}\mbf{E}\times\mbf{H} \\
\frac{1}{c}\mbf{E}\times\mbf{H} & -\mbf{E}\wedge\mbf{D}-\mbf{H}\wedge\mbf{B}+\frac{1}{2}\left(\mbf{E}\cdot\mbf{D}+\mbf{H}\cdot\mbf{B}\right)\mbf{I}
\end{array}\right)\label{eq:AbrEMEMT}\\
T^{\mu\nu}_{\mrm{mat, Abr}}&=&\left( \begin{array}{cc}
\rho_0 (c^2+\epsilon_i) & \rho_0 c \mbf{v}\\
\rho_0 c\mbf{v} & \rho_0\mbf{v}\wedge\mbf{v} + \phi\mbf{I}
\end{array}\right)\label{eq:AbrMat}\\
T^{\mu\nu}_{\mrm{EM, Mink}}&=&\left( \begin{array}{cc}
\frac{1}{2}\left(\mbf{E}\cdot\mbf{D}+\mbf{H}\cdot\mbf{B}\right) & \frac{1}{c}\mbf{E}\times\mbf{H} \\
c\,\mbf{D}\times\mbf{B} & -\mbf{E}\wedge\mbf{D}-\mbf{H}\wedge\mbf{B}+\frac{1}{2}\left(\mbf{E}\cdot\mbf{D}+\mbf{H}\cdot\mbf{B}\right)\mbf{I}
\end{array}\right)\label{eq:MinkEMEMT}\\
T^{\mu\nu}_{\mrm{mat, Mink}}&=&\left( \begin{array}{cc}
\rho_0 (c^2+\epsilon_i) & \rho_0 c \mbf{v}\\
\rho_0 c\mbf{v}-c\,\mbf{D}\times\mbf{B}+\frac{1}{c}\mbf{E}\times\mbf{H} & \rho_0\mbf{v}\wedge\mbf{v} + \phi\mbf{I}
\end{array}\right)\label{eq:MinkMat}.
\end{eqnarray}
\end{widetext}

In this representation, the terms in the Minkowski material counterpart giving rise to the Abraham force may clearly be seen in Eq. (\ref{eq:MinkMat}).
In contrast, the field-related component of the Abraham material momentum tensor arises from an implicit dependence of $\mbf{v}$ on \textbf{E}, \textbf{B}, \textbf{D}, and \textbf{H}. 
Also note that although the Minkowski material momentum density is frequently very small or zero, this does not reflect the behaviour of the particles of the dielectric medium, whose velocity is related only to the term $\rho_0 c \mbf{v}$.

Now consider a wave packet of total momentum $p$ and volume $V$ incident upon a dielectric slab of group refractive index $n_g$. The velocity of the wave packet slows to $c/n_g$ on entering the slab, and its volume reduces to $V/n_g$. The total momentum must remain constant at $p$, and hence the total momentum density increases from $\mbf{g}_f$ in free space to $n_g\mbf{g}_f$ within the material. However, under the Abraham tensor the electromagnetic momentum density within the material is only $\mbf{g}_f$ and hence it must be accompanied by a material momentum of $(n_g-1)\mbf{g}_f$. Identifying this with the corresponding term in (\ref{eq:AbrMat}), and substituting for $\mbf{g}_f$ the free space momentum density $\mbf{E}\times\mbf{H}/c^2$, we obtain
\begin{equation}
\mbf{g}_\mrm{mat,Abr}=\rho_0\mbf{v}=\left(n_g-1\right)\frac{\mbf{E}\times\mbf{H}}{c^2}.
\end{equation}
This expression holds for both dispersive and nondispersive media. It is trivial to show that the same expression for $\mbf{v}$ is obtained under the Minkowski tensor pair.

This result differs from the expression obtained by \textcite{jones} and \textcite{jones4}, who argued that an electromagnetic wave propagating in a dielectric medium at velocity $c/n_g$ with momentum density $\mbf{g}_m$ will give rise to a momentum flux $\mbf{g}_m/n_g$. Based on an observed momentum flux of $n_\phi \mbf{g}_f$, where $\mbf{g}_f$ is the momentum density in free space, they concluded the total momentum density associated with the wave within the dielectric must be given by $\mbf{g}_m=n_g n_\phi \mbf{g}_f$. This yields a material counterpart to be added to the Abraham momentum density of
\begin{equation}
\mbf{g}_\mrm{mat,Jones}=\left(n_\phi n_g-1\right)\frac{\mbf{E}\times\mbf{H}}{c^2}\label{matjones}
\end{equation}
which simplifies to
\begin{equation}
\mbf{D}\times\mbf{B} - \mbf{E}\times\mbf{H}/c^2\label{abrterm}
\end{equation}
in nondispersive media and generates the Abraham force (\ref{eq:abrterm}). However, this result is at odds with that of \textcite{walker}, who demonstrated that the Abraham electromagnetic momentum density is correct, requiring that Eq.~(\ref{matjones}) or (\ref{abrterm}) instead be deducted from the Minkowski momentum density, as seen in our Minkowski material tensor (\ref{eq:MinkMat}).

The error here is in the treatment of momentum flux. While the expression $\mbf{g}_m/n_g$ will hold for a single wave propagating in isolation through the dielectric medium, it breaks down for superpositions such as the incident and reflected beams present here, and $\partial_j\,\mbf{T}^{ij}$ must be used instead. 
The incident beam is associated with a medium particle velocity of $\mbf{v}$ and the reflected beam with a particle velocity of $-\mbf{v}$. While the accompanying fields interfere, the velocities simply add, and the medium adjacent to the mirror remains stationary. The momentum flux is therefore dependent only on field-related terms. Unlike $\mbf{g}_m/n_g$, $\partial_j\,\mbf{T}^{ij}$ is insensitive to the choice of Abraham or Minkowski momentum density.
Taking proper heed of the effect on the magnetic field of time-varying currents and charges induced within the medium, we obtain via Eq.~(\ref{eq:boundary}) 
a flux of $n_\phi \mbf{g}_f$, which is in accordance with experiment.
As this result
is independent of the choice of Minkowski or Abraham electromagnetic energy--momentum tensor, we are free to assign (\ref{abrterm}) to the Minkowski tensor in agreement with \textcite{walker}.


We also note a further advantage to our expression for momentum density: Consider a wave packet of total momentum $p$ and unit volume entering the medium. Under the total energy momentum tensor approach given here,
the volume of the wave packet falls to $1/n_g$, its total momentum density rises by a factor of $n_g$, and momentum is conserved. Under Jones and Richards' approach the 
total momentum density rises by a factor of $n_\phi n_g$, giving a momentum associated with the wave packet of $n_\phi p$. Conservation of momentum is usually 
restored by positing a transfer of momentum $-(n_\phi-1)p$ to the medium boundary \cite[see, for example,][]{loudon3}. Nevertheless, in steady state experiments such as \textcite{jones2} we might 
expect this momentum to be conducted to the mirror via pressure effects within the 
medium, reducing momentum transfer to the mirror at equilibrium from $n_\phi p$ to just
$p$. 
This reduction is not observed, and this supports our approach over
that of Jones and Richards.




\subsection{Forces Acting On a Subsytem\label{subsystem}}

While it is always possible to return to first principles, in the manner of \textcite{gordon}, and individually evaluate all the forces acting in a situation, this is frequently a far from simple task. However, where appropriate expressions for the total energy--momentum tensor are known, force and torque densities for a specific subsystem may be calculated directly. The means by which they are calculated will depend upon whether the subsystem is spatially or qualitatively distinguished from the system as a whole.


We use the covariant generalisation of angular momentum density
\begin{equation}
M^{\alpha\beta\gamma}=T^{\alpha\beta}x^\gamma-T^{\alpha\gamma}x^\beta \label{eq:defineM}
\end{equation}
given in \textcite{jackson}, where $x$ is the 4-vector position co-ordinate relative to the point about which torque is determined. $M^{0jk}$ is the 3x3 antisymmetric angular momentum tensor whose three independent elements map onto the angular momentum density pseudovector $\mbf{l}$ via
\begin{equation}
l_i=\frac{1}{2}\,\varepsilon_{ijk}M^{0jk}
\end{equation}
where $\varepsilon$ is the Levi--Civit\`{a} symbol. The elements $M^{0j0}=-M^{00k}$ represent the puzzling yet orthogonally necessary temporal components of angular momentum, and $M^{i\beta\gamma}$ describes angular momentum flux, analogous to $T^{i\beta}$ in the total energy--momentum tensor. From (\ref{eq:defineM}), (\ref{nodiv}), and (\ref{symmetry}), it follows that
\begin{equation}
\partial_\alpha\,M^{\alpha\beta\gamma}=0,
\end{equation}
which corresponds to conservation of angular momentum.

The torque density tensor will be denoted $t^{jk}$, and 
maps onto the torque pseudovector via $t_i=\frac{1}{2}\varepsilon_{ijk}t^{jk}$.

\subsubsection{Qualitatively Distinguished\label{qualdis}}

Let the system be divided into two parts, $A$ and $B$, which are qualitatively different but occupy the same region of physical space. We wish to calculate the momentum transfer from $A$ to $B$. For example, $B$ might be a material component, such as a dielectric medium, and $A$ the electromagnetic field. We also similarly divide the linear momentum density, angular momentum tensor, and stress tensor into parts $A$ and $B$. Into $\mbf{g}_B$, $M_B$ and $\mbf{T}_B$ we place all terms relating to the physical momentum of the material component, as it is the force and torque on the material that we wish to evaluate. The remainder of the terms are placed into $\mbf{g}_A$, $M_A$ and $\mbf{T}_A$. The force and torque densities on component $B$ are then evaluated using
\begin{eqnarray}
\mbf{f}^{i}_B &=& - \partial_t \, \mbf{g}_{A}^{i} - \partial_{j} \, \mbf{T}_A^{ij}\label{eq:fdens}\\
t^{jk}_B&=& - \partial_t \,M^{0jk}_A + \partial_i \,M^{ijk}.\label{eq:tdens}
\end{eqnarray}
These are momentum balance equations, with the rate of momentum and torque transfer into system $B$ equalling the rate of loss from system $A$.

Integrating over the volume of the subsystem then gives the total force or torque exerted on $B$ by $A$ as a result of coupling between the two subsystems, for example absorption of momentum from an electromagnetic field.
\begin{equation}
\mbf{F}^i_B=\int\mrm{d}V \, \mbf{f}^i_B,\quad \tau^{jk}_B=\int\mrm{d}V \, t_B^{jk}.\label{eq:volforce}
\end{equation}
$\mbf{F}$ and $\tau$ represent the total force and torque respectively.

The introduction of a complex refractive index allows the modelling of electromagnetic wave propagation through media demonstrating loss or gain, and it is the resulting dependency of the magnitudes of the electric and magnetic fields on position and time that gives rise to a non-zero time average for the force or torque of (\ref{eq:fdens}) and (\ref{eq:tdens}). Of course, by the Kramers--Kronig relations any absorptive material medium is also necessarily dispersive. This may be modelled by describing a separate energy--momentum tensor with appropriate dielectric and magnetic constants for each mode of the electromagnetic wave, and expressing the refractive index as a complex function of the wave number, $n(k)$.


For near-monochromatic waves in the steady state, dispersion may frequently be neglected and the use of a single electromagnetic energy--momentum tensor with the material counterpart tensor may suffice.

\subsubsection{Spatially Distinguished}

Now suppose we are interested in momentum flux into or out of a particular region of space. We denote this region $B$ and the outside region $A$. By Gauss's Law, the rate of change of momentum within region $B$ is dependent upon the rate of flux through the boundary. Transforming Eq.~(\ref{eq:volforce}) into surface integrals, we obtain
\begin{eqnarray}
\mbf{F}^i_B&=&-\oint\mrm{d}S_j\, \mbf{T}^{ij}_{A,\mrm{in}}-\oint\mrm{d}S_j\, \mbf{T}^{ij}_{B,\mrm{out}}\label{eq:boundary}, \\
\tau^{jk}_B&=& \oint\mrm{d}S_i\, M^{ijk}_{A,\mrm{in}} + \oint\mrm{d}S_i\, M^{ijk}_{B,\mrm{out}}.
\end{eqnarray}
The normal to the boundary is taken as pointing from $B$ to $A$. The first integral in each equation evaluates flux from $A$ to $B$, and only terms inbound to region $B$ should be retained. Similarly, the second integral evaluates flux from $B$ to $A$ and only outbound terms should be retained.
It is not required that regions $A$ and $B$ be materially identical. Evaluation of these surface integrals may frequently be simplified by judicious application of Gauss's law.


Note that the form of the momentum entering area $B$ --- mechanical or electromagnetic --- is not specified.

\subsubsection{General Subsystem}

In more general circumstances, for example momentum transfer to a solid dielectric block immersed in a dielectric fluid and exposed to an electromagnetic field, both spatial and qualitative considerations may apply. The block may receive momentum through its boundaries as a result of the external medium (e.g. pressure) and the external electromagnetic field (e.g. reflection), and within its bulk as a result of coupling between matter and the electromagnetic field within the block (e.g. absorption). We denote the region containing the block region $B$, and the external region $A$.

As in Sec.~\ref{qualdis}, we divide the linear momentum density, angular momentum tensor, and stress tensor within the block into material components, denoted $\mbf{g}_{B,\mrm{mat}}$, $M_{B,\mrm{mat}}$, and $\mbf{T}_{B,\mrm{mat}}$, and electromagnetic components, denoted $\mbf{g}_{B,\mrm{EM}}$, $M_{B,\mrm{EM}}$, and $\mbf{T}_{B,\mrm{EM}}$, purely dependent upon whether they relate to the outcome we wish to determine --- the physical momentum of the dielectric block. We then label the region external to the block region $A$, and similarly divide its stress tensor into material and other components, except that here, a term is considered material if it will transfer momentum into the material component of section $B$ on encountering the boundary.
The equations for the force and torque acting on the dielectric block then become
\begin{eqnarray}
\mbf{F}^i_{B,\mrm{mat}}&=&- \int\mrm{d}V \,\left( \partial_t \, \mbf{g}_{B,\mrm{EM}}^{i} + \partial_{j} \, \mbf{T}_{B,\mrm{EM}}^{ij}\right) \nonumber\\
&&-\oint\mrm{d}S_j\,\left(\mbf{T}^{ij}_{A,\mrm{mat,in}}+\mbf{T}^{ij}_{B,\mrm{mat,out}}\right)\label{eq:fullforce}\\
\tau^{jk}_{B,\mrm{mat}}&=&- \int\mrm{d}V \,\left( \partial_t \, M_{B,\mrm{EM}}^{0jk} - \partial_{i} \, M_{B,\mrm{EM}}^{ijk}\right)\nonumber\\
&&+\oint\mrm{d}S_i\,\left(M^{ijk}_{A,\mrm{mat,in}}+M^{ijk}_{B,\mrm{mat,out}}\right).\label{eq:fulltorque}
\end{eqnarray}
Further subdivision of field-related tensors into the different electromagnetic modes may be required for the modelling of dispersive media, as indicated in Sec.~\ref{qualdis}.


The labelling of terms in region $A$ as ``material'' should not be taken literally. For example, if the block in region $B$ is perfectly reflective, then all electromagnetic waves in region $A$ would be classified as ``material'' as they would impart their momentum to the material component of region $B$.
The designation of ``material'' is functional and context-dependent, and should not be taken as proposing a clear and unique subdivision in the manner of the Abraham or Minkowski approaches.

At this point we will be well served by examining some examples within the literature.

\subsection{Examples\label{examples}}

\subsubsection{Reflection of Light Off a Mirror\label{ex:mirror}} 

Consider the reflection of light off a perfect mirror suspended within a dielectric. We wish to evaluate the momentum transferred to the mirror, as in \textcite{jones} or \textcite{jones2}. This takes place only at the surface of the mirror, and therefore we can use the surface integral (\ref{eq:boundary}). For $\mbf{T}_A$ we take the total stress tensor within the dielectric. We may discard the term in $\mbf{T}_B$ as there will be no momentum flux from the mirror to the dielectric.
As noted in Sec.~\ref{sec:atotal}, the medium adjacent to the mirror remains stationary because the velocities induced by the incident and reflected beams sum to zero. Momentum is therefore transferred by the electromagnetic fields alone.

Historically the Minkowski electromagnetic energy--momentum tensor in isolation has also been used here with great success. Although the assumed momentum density differs from that yielded by the total energy--momentum tensor, the simultaneous
identification of the momentum flux as $\mbf{g}/n_g$ instead of $\partial_j\,\mbf{T}^{ij}$
yields the correct overall result, up to a factor of $n_g/n_\phi$.
This works because the Minkowski material momentum density is typically numerically very close to zero, and hence the total momentum density may be reasonably approximated by the electromagnetic momentum density $\mbf{D}\times\mbf{B}$. That the total momentum flux is given by $\mbf{g}/n_g$ then follows from Gauss's Law. The result is satisfactory when 
the error introduced by this approximation is small.


\subsubsection{Transmission of Light Through a Nonreflective Dielectric Block\label{nonreflec}}

Now consider a beam of light incident upon a block of perfectly nonreflective dielectric of higher refractive index than the surrounding medium. \textcite{balazs} considers the transmission of a wave packet through a glass rod with anti-reflection coatings on both ends. If we are interested in the behaviour of the surface, then 
surface terms must be taken into account using (\ref{eq:fullforce}), but if we are only interested in the motion of the centre of mass of the block, then we can neglect the surface forces provided the time average of the integral (\ref{eq:boundary}) goes to zero over a period much less than the sensitivity of our apparatus.

For a laser beam traversing a dielectric block, we may eliminate the field-related terms of the integral if the electromagnetic fields where the beam enters and leaves the block are identical up to a phase. At its simplest this requires zero reflection, zero absorption, and that the two surfaces be parallel. We may eliminate the remainder of the terms by choosing the surrounding medium to be vacuum. The behaviour of the centre of mass of the block may then be determined using (\ref{eq:volforce}). For $n_\phi>1$, the block accelerates away from the beam source while it is traversed
by the leading edge of the beam, then continues to travel away from the source at constant velocity while the beam is turned on. When the beam is turned off, traversal of the trailing edge restores the block to rest. The use of (\ref{eq:volforce}) can
also readily be extended to include absorption, giving rise to an additional ongoing acceleration.

\subsubsection{Transmission and Reflection Off a Dielectric Block}

The most general case to consider is a partially reflective dielectric block suspended within an arbitrary dielectric medium. In the absence of simplifying assumptions we must employ the full expression (\ref{eq:fullforce}), with both a surface and a volume term. We may compare this with the work of \textcite{mansuripur}, who calculated the forces on a dielectric using the Lorentz force. Integration of the Lorentz force density over the volume of the dielectric gives a volume term, whereas reflection of a portion of the incoming radiation gives rise to a surface term. Although Mansuripur examined pressure and tension arising within the dielectric block as a result of the incident radiation, contributions of these effects to the net surface force, as in \textcite{gordon}, are not discussed. These forces make no contribution to the motion of the centre of mass, but will affect the behaviour of the surface if the block is treated as flexible.

Mansuripur obtained an expression for the electromagnetic momentum density in a dielectric medium from the rate of transfer of momentum into the medium both as electromagnetic fields and via the Lorentz force, but excluded momentum transferred to the surface via reflection, which he considered non-electromagnetic. However, this merely represents a further scheme by which to subdivide the total energy--momentum tensor, and when electromagnetic radiation passes from one medium into another, both bulk and surface forces will be present. Momentum flux through the boundary under this scheme is therefore indistinguishable from the combined electromagnetic and material tensors of either Abraham or Minkowski. 
How Mansuripur's definition of the electromagnetic momentum density extends to moving media has yet to be established.

In general, momentum transfer to the surface of a dielectric block arises both from interactions with the external environment and from pressure effects within the block itself. This may cause regions of relative compression or rarefaction close to the surface, depending on whether the surface force is of the same or opposite sign to the Lorentz force which acts on the bulk of the block. For dielectric fluids, a flow may result. When Eq.~(\ref{eq:fullforce}) is used, these effects are taken into account via the material components of the total energy--momentum tensor.
Such effects are seen in the experiment of \textcite{ashkin} discussed in Sec.~\ref{expts:ashkin}, where the surface of the water bulges towards the source of the laser beam, although the Lorentz force within the body of the liquid acts in the opposite direction.



\subsection{Discussion; Simplifying Assumptions\label{discussion}}

It is frequently possible to make simplifying assumptions. We have already considered
a perfectly reflecting mirror, for which no volume term arises, and a perfectly transmissive dielectric in vacuum, for which the surface term has no effect on the motion of the centre of mass.




Also, it is often possible to use the Abraham or Minkowski energy--momentum tensors in the traditional manner. The Abraham tensor pair are well suited to problems in highly transmissive dielectrics, such as Example~\ref{nonreflec}, as momentum transfer to the material component may be identified with the Lorentz force \cite{stallinga2}. The Minkowski electromagnetic energy--momentum 
density is frequently used to calculate momentum transfer across a boundary, as discussed in Sec.~\ref{ex:mirror}.

It is important to note that electrostriction, magnetostriction, dispersion, and acoustic waves are rarely treated in the literature, though with some exceptions \cite{mikura,nelson,stallinga,garrison}. When these effects are significant, an appropriate material energy--momentum tensor must always be used.

Rigorous treatment using a Lorentz force approach \cite{loudon4,mansuripur,mansuripur2} is also a viable option under any circumstances, provided sufficient care is taken over
field/charge interactions at the surface \cite{mansuripur}, and may frequently be the simpler approach. Where the behaviour of the surface is important, for example in the behaviour of dielectric liquids \cite{ashkin}, stress- or pressure-related effects must also be considered.

\section{Anomalous Effects?\label{anomeffects}}

Over the years, a large number of tensor-dependent effects have been proposed, by which it has been suggested that it might be possible to distinguish between the different tensor formulations. These tend to arise either from inadequately considering the role of the material energy--momentum tensor, or from inappropriately treating the dielectric medium as a rigid body.

Failure to adequately consider the role of the material energy--momentum tensor may lead to claims of experimental discrimination between the different energy--momentum tensors, as in \textcite{walker}, and hypothetical thrusters based on non-conservation of momentum \cite{brito}.

Perhaps more important is the error of treating the dielectric as a rigid body, an error which recurs frequently within the literature \cite{balazs,arnaud,arnaud2,brevik3,padgett} and also leads to inappropriate predictions of unusual physical behaviour.

\subsection{Rigid Bodies and the Optical Tractor Beam\label{rigidity}}

Consider a pulsed electromagnetic wave of momentum $p$ entering a material body of refractive index $n$. If we use the Minkowski electromagnetic energy--momentum tensor, on entering the body the momentum of the wave is increased to $np$ and hence by conservation of linear momentum, the body acquires a momentum of $-(n-1)p$, setting it in motion towards the source of the wave. When the pulse exits the body, the momentum of the material is restored to zero and the body returns to a stationary state. In contrast, under the Abraham tensor the momentum of the pulse decreases and the body must begin moving {away} from the source of the wave.
\textcite{jones4}
discussed such behaviour in the context of an Einstein box --- a rigid box in which photons are emitted from one end and absorbed by the other, typically used as a thought experiment to demonstrate mass--energy equivalency.

Because the body is treated as rigid and hence unable to deform, momentum is transferred to the body as a whole, propagating across it instantaneously. In reality, the transfer of momentum must take place at a finite velocity which cannot exceed $c$ and will therefore either be much less than, or of comparable magnitude to, the speed of light in the medium, $c/n$. Because of this, significant compressive and expansive effects must take place, which are neglected if the body is treated as rigid.


This approach also mis-handles momentum associated with the material body under the Minkowski tensor. As seen in (\ref{eq:MinkMat}), the Minkowski material momentum density includes a term $(1/c)\mbf{E}\times\mbf{H}-c\,\mbf{D}\times\mbf{B}$, 
which causes
the material momentum density as a whole to be negative. However, this term is not associated with the movement of material particles described by $\mbf{v}$, and is therefore not related to the motion of material particles within the block. While it is necessary for the material tensor momentum density to be negative in order to offset the increase in momentum density in the Minkowski electromagnetic tensor, not all of this momentum density is associated with the particles of the block. In fact, the behaviour of these \textbf{v}-dependent terms is identical to that under the Abraham tensor and the block gains positive momentum, although the material momentum density as a whole becomes negative under the Minkowski scheme.

When these effects are treated correctly, we recover an approach equivalent to that of \textcite{gordon},
as discussed in Sec.~\ref{sec:gordon}, and the traversal of the light pulse now generates a distortion in the boundary of the medium.
When the leading edge of the pulse enters the medium, the boundary bulges outwards, as demonstrated by the experiments of
\textcite{ashkin}, and
a similar outward bulge is produced when the leading edge
exits the medium.
The traversal of the trailing edge of the pulse gives rise to a restoring impulse in each case.
Meanwhile, 
the unopposed action of the Lorentz force within the bulk of the medium
gives rise to the motion of the body as a whole.
The resulting behaviour is identical under both the Abraham and Minkowski tensors.

As discussed in Sec.~\ref{mompseudomom}, \textcite{balazs} demonstrated that a perfectly transmitting dielectric of higher refractive index than its surroundings will advance in the direction of propagation of the wave packet, using arguments relating to the movement of the centre of mass of the system.
However he then erroneously applies the rigid body argument given above, 
and concludes
against the Minkowski electromagnetic energy--momentum tensor. In reality, as we have seen in Sec.~\ref{nonreflec}, both choices when used correctly will yield the same 
Lorentz forces acting on the body of the rod, accompanied by any surface effects due to pressures within the rod as per \textcite{gordon} and \textcite{peierls}, because regardless of whether we choose the Minkowski or Abraham representation, the total energy--momentum tensor remains unchanged.

We note that in a real medium, coupling between adjacent surface elements (for example, surface tension in a liquid or lattice forces in a solid) may cause disturbances generated by surface forces to propagate radially outwards. Thus in an absorptive medium, traversal of a radiation pulse may give rise to acoustic (matter) waves within the substance of the medium, but once again the predicted behaviour will be independent of the choice of energy--momentum tensors used. \textcite{peierls}
showed that such photon/phonon coupling may also take place where the edges of a finite beam lie within a dielectric.

\section{Recent Work\label{recentwork}}

As mentioned in the introduction, current research on the electromagnetic energy--momentum tensor relates largely to practical applications, or to the extension of existing knowledge into new domains, though there are still many who appear unaware that the Abraham--Minkowski controversy has been resolved. We present and discuss some of the more recent papers, to illustrate the present state of research in the field.

\subsection{Extension of Existing Theory and Practical Applications}

Obviously, any theoretical result requires verification. The ability of combined electromagnetic and material energy--momentum tensors constrained by conservation of linear and angular momentum to explain historical experiments such as those discussed in this paper is well established. However, continued testing of such theories over an increasingly broad range of domains is an essential part of the process of physics research. In this regard, we refer to the proposal of
\textcite{antoci2} to demonstrate 
the Abraham force at optical frequencies, 
and the work of
\textcite{brito} seeking violation of conservation of momentum, though his
reported positive 
results
fare poorly under independent scrutiny \cite{nasa}.

Extending existing classical theory,
\textcite{mansuripur}
directly derived the force of electromagnetic radiation on a material medium from the Lorentz force law, an approach which makes no pre-emptive assumptions regarding the value of the electromagnetic momentum. He reproduced the beam edge effects first discussed by
\textcite{gordon},
but with the discovery that this force may be expansive, rather than constrictive, for $p$-polarised light, an effect 
which awaits
experimental verification. He also deduced that anti-reflective coatings experience a force which attempts to peel them off their substrate. Like many others, in the course of this paper he derived a unique ``correct'' expression for the electromagnetic momentum density. There are almost as many such expressions as there are authors, and if we restrict physics to dealing only with hypotheses which may be tested by experiment, then the question of the correct expression
is instead one of ontology.
However, as physical predictions are independent of the expression chosen (see Sec.~\ref{proofsequiv}), this in no way detracts from the significance of his other findings.

Other recent work includes the experiment of
\textcite{campbell}
to determine whether, when a photon is absorbed by an atom within an atomic cloud, it should be considered as departing the medium, and therefore only transfer the momentum of $\hbar k$ to the atom, in accordance with
\textcite{hensley}, or whether it transfers the Minkowski value $n\hbar k$, leaving behind a disturbance in the medium, as proposed by
\textcite{haugan}.
The latter case was determined to be correct, with significance for the construction and operation of atom interferometers.

Using quantum theory,
\textcite{garrison} developed operators for the momentum of a photon in a dielectric medium. Employing a quantisation scheme proposed by
\textcite{milonni},
they arrived at a situation they recognise to be directly analogous to the Abraham--Minkowski controversy. Their survey of the literature included the recognition, which they attributed to
\textcite{brevik3}, that decomposition of the total energy--momentum tensor into electromagnetic and material components is arbitrary. However, due to the inconsistent nature of the available literature, this was stated as simply one possible position of many, rather than an obligate consequence of momentum conservation. \citeauthor{garrison} proceeded to develop three momentum operators --- Abraham and Minkowski momentum operators, corresponding to classical momenta calculated utilising the isolated electromagnetic energy--momentum tensors of Abraham and Minkowski, and a canonical momentum operator, which is the generator of translations, and relates to the total energy--momentum tensor. They reanalysed the results of
\textcite{jones2}, and demonstrated that the experiment favours the canonical momentum operator over both the Abraham and Minkowski operators. This is to be expected, as no operators were employed which related to either the corresponding material momentum flux or the Maxwell stress tensor, and hence only the canonical momentum operator accounts for the total flux of momentum within the system. Their analysis is novel, in that it demonstrates that the results of Jones and Leslie are of sufficient accuracy to distinguish between the total energy--momentum tensor, and the electromagnetic energy--momentum tensor of Minkowski in isolation. By doing so, they therefore provide additional experimental verification that the Minkowski tensor, like the Abraham tensor, is incomplete without a material counterpart. 
For any real medium the Minkowski material counterpart tensor will yield not only the Abraham term, but also terms giving rise to acoustic transients \cite{peierls} and a dispersion-related coupling between the electromagnetic wave and the material medium \cite{garrison,stallinga,nelson}. 
This latter effect
is responsible for the 
observed discrepancy between the canonical momentum and the Minkowski electromagnetic momentum. 

Garrison and Chiao concluded that the canonical (total) momentum is most applicable for interpreting experiments, and question whether the use of Abraham and Minkowski operators can ever be appropriate. However, they weaken their conclusion by citing the experiment of
\textcite[and Sec.~\ref{expts:walker} above]{walker}
as uniquely selecting the Abraham tensor, despite having earlier cited
\textcite{obukhov}, who indirectly disprove this assertion. They also present the arguments of
\textcite{lai} as similarly 
favouring the Abraham tensor, though these have been refuted by 
\textcite{brevik4}, as discussed in Sec.~\ref{arguments}. Garrison and Chiao do not examine the experiment of Walker \emph{et al.} or the suggestion of 
\textcite{lai3} 
in their quantum-mechanical formalism, which is a pity, since they might then have concluded that once again, it is conservation of the canonical momentum which governs the experiments' behaviour.

\textcite{loudon3} 
also investigate momentum transfer from a quantum perspective, looking at the absorption of photons by charge carriers in a semi-conductor. They found that momentum transfer can be divided into three separate components: momentum transfer to the charge carriers, given by the Minkowski value $n_\phi \hbar k$ and in agreement with previous experiment \cite{gibson}, to the bulk of the semi-conductor, given by the Abraham value $\hbar k/n_{\mathrm{g}}c$, and to the surface of the dielectric, yielding overall conservation of momentum. A classical treatment by
\textcite{mansuripur2} 
again results in transfer of the Minkowski value of momentum to the charge carriers, with the net action on the host semiconductor maintaining global conservation of momentum.

\textcite{feigel} 
has looked at the transfer of momentum between matter and the electromagnetic field on the quantum scale, and predicted the intriguing phenomenon of fluid flow powered by quantum vacuum fluctuations which is dependent upon the high-frequency cutoff of quantum field theory. 

Several have found the electromagnetic energy--momentum tensor important in extension of the Casimir effect from vacuum-separated plates to those separated by a dielectric matierial. 
\textcite{raabe} related the importance of recognising that the Minkowski tensor on its own is incomplete, and
\textcite{stallinga} 
suggested that consideration of the total momentum flux density is most appropriate. 
Stallinga also demonstrated how a classical technique utilising an artificial ``auxiliary field'' $\mathbf{F}$ may be useful in modelling dissipative systems.

Of interest for different reasons is the somewhat older work of 
\textcite{cole,cole2} 
in representing the behaviour of massless photons in a material medium as that of equivalent massive particles in vacuum.
Due to a transformation effectively eliminating the material medium, all momentum is contained within the electromagnetic energy--momentum tensor, and in this special case, the criteria of von Laue and M\o{}ller (see Sec.~\ref{arguments}) are fulfilled. In terms of the Abraham--Minkowski controversy, Cole
is in effect choosing an energy--momentum tensor in which all active terms are placed within the electromagnetic tensor and the material energy--momentum tensor is always null. More recently, 
\textcite{antoci3} 
have drawn attention to a related treatment of the constitutive equation of electromagnetism by 
\textcite{gordon2}, 
valid for a medium homogeneous and isotropic in its local rest frame.

A similar approach is also of practical 
relevance in modelling optical trapping. Since all forces acting due to transfer of momentum from fields and media must be considered, and the trapped particle is typically suspended in a dielectric medium, 
a separation of the field and material components of the momentum flux that allows the surrounding medium to be ignored can simplify calculation of the optical force. This separation requires that the surrounding medium carries none of the momentum,
a condition which is adequately met by the Minkowski tensor provided both dispersion and the Abraham force 
are negligible, the fluid medium has had time to reach equilibrium, and behaviours 
such as electrostriction and magnetostriction, which are not supported by the electromagnetic component of this tensor, may be neglected.
The trapped particle can then be modelled as a dielectric particle of refractive index $n_\mathrm{relative} = n_\mathrm{particle}/n_\mathrm{medium}$ in free space, with the beam delivering a momentum flux density of $n_\mathrm{medium}\mathbf{S}/c$. 
This method is best suited to monochromatic light or nondispersive media, where $n$ may be treated as a constant.

For a rigid particle, the optical forces can be calculated from the incoming and outgoing momentum fluxes~\cite{nieminen2004d}. For a deformable particle, for example in an optical stretcher~\cite{guck2001}, one needs to include the mechanical elastic forces that resist the field or medium forces acting to deform the particle, as well as the field or medium forces themselves. While it may be simplest to assume the Minkowski momentum,
the alternative tensors are nevertheless physically equivalent. Statements which are based on the interpretation of a single tensor, for example suggesting that the deforming force arises because the momentum of a photon in a medium is $n\hbar k_0$, compared to $\hbar k_0$ in vacuum, are therefore unwise.

\subsection{Ongoing Controversy}

The work of 
\textcite{penfield}, \textcite{degroot}, \textcite{gordon}, \textcite{mikura}, \textcite{kranys,kranys3,kranys2}, \textcite{maugin}, and \textcite{schwarz} 
effectively demonstrated that 
division of the total energy--momentum tensor into electromagnetic and material parts is 
effectively
arbitrary. Nevertheless, many continue to attempt to identify a superior expression for the electromagnetic energy--momentum tensor. Often they are aware of the work of Penfield and Haus, and de Groot and Suttorp, and one must therefore suppose that these works, perhaps on account of complexity or lack of sufficient emphasis, are proving ineffective in conveying that such a distinction is at the very least 
untestable. For Gordon, one may point to the limited scope to which his treatment may be applied, and for Mikura, Krany\v{s} and Maugin the problem appears to be 
insufficient publicity.

Whatever the cause, 
papers continue to appear which attempt to demonstrate a uniquely privileged
form for the electromagnetic
energy--momentum tensor
\cite[see, for example,][]{antoci,obukhov,labardi,mansuripur,padgett,scalora}.
Padgett proposed an interesting experiment in the optical domain, 
in which a laser beam carrying orbital angular momentum traverses, and transfers angular momentum to, a glass disc. This experiment is in many ways analogous to that of \textcite[and Sec.~\ref{expts:ashkin} above]{ashkin}. In Ashkin and Dziedzic's experiment, pressure effects within the liquid were responsible for restoring equivalence between the Abraham and Minkowski interpretations. In Padgett's experiment, a similar role will be played by shear forces within the disc.


The works cited above are far from alone in their continued fascination with the Abraham--Minkowski controversy. The topic frequently continues to be a point of discussion at conferences, and the occasional physics colloquium \cite{colloquium}, though awareness is gradually increasing that the distinction is of no functional significance,
and hence the choice of energy--momentum tensor pair is essentially a question of personal \ae{}sthetics.

\section{Conclusion}

The original Abraham--Minkowski controversy, over the preferred form of the electromagnetic energy--momentum tensor in a dielectric medium, has been resolved by the recognition that division of the total energy--momentum tensor into electromagnetic and material components is arbitrary \cite{penfield,degroot}. Hence the Minkowski electromagnetic energy--momentum tensor, like the Abraham tensor, has a (frequently unacknowledged) material counterpart \cite{israel,obukhov}, and the sum of these components yields the same total energy--momentum tensor as in the Abraham approach.

On these grounds, all choices for the electromagnetic energy--momentum tensor are equally valid and will produce the same predicted physical results, as has been demonstrated for a wide range of specific examples by
\textcite{mikura}, \textcite{kranys,kranys2,kranys3}, and \textcite{maugin}.
Nevertheless, awareness of the resolution of the original controversy remains patchy, largely due to the fragmentary nature of the literature on the subject.

In this paper, we have reviewed the controversy from its initial formulation to the present day, incorporating discussion of key experiments believed at one time or another to have bearing on the relative truths of the Abraham and Minkowski models. We have discussed the realisation that any electromagnetic energy--momentum tensor must always be accompanied by a counterpart material energy--momentum tensor, and that the division of the total energy--momentum tensor into these two components is entirely arbitrary.

A consensus has yet to be reached on the ultimate form of the total energy--momentum tensor, but the limitation here arises not from classical electromagnetism but from materials science, and we believe there is scope for further work in explicitly developing the theory of dispersive media, a subject touched upon in Sec.~\ref{qualdis}.



We believe a wider appreciation of the relationship between electromagnetic and material energy--momentum tensors could greatly benefit those working in this field, and with this colloquium we hope to assist in bringing that wider appreciation into existence.

\begin{acknowledgments}
The authors would like to thank The Royal Society, Professor A. Ashkin, the American Physical Society, and the National Research Council of Canada for permission to reproduce the figures accompanying this paper. The work of R.N.C.P was partly supported by an Endeavour International Postgraduate Research Scholarship funded by the Australian Government Department of Education, Science and Technology.
\end{acknowledgments}

\end{document}